\documentclass{svjour3}                     
\smartqed  
\usepackage{graphicx}
\usepackage{bm} 
\usepackage{color}

\newcommand{\DART}{DART\textsuperscript{\rm\textregistered}}

\newcommand{\bl}[0]{\color{blue}}
\journalname{Natural Hazards}

\begin{document}

\title{Evaluating Effectiveness of \DART\ Buoy Networks}

\author{Donald~B.~Percival        \and
        Donald~W.~Denbo \and
        Edison~Gica \and
        Paul~Y.~Huang \and
        Harold~O.~Mofjeld \and
        Michael~C.~Spillane \and
        Vasily~V.~Titov
}

\institute{Donald B.~Percival \at
            Applied Physics Laboratory,
            Box 355640,
            University of Washington,
            Seattle, WA  98195-5640
            USA \\
              \email{dbp@apl.washington.edu}           
           \and
           Donald B.~Percival \at
           Department of Statistics,
           Box 354322,
           University of Washington,
           Seattle, WA  98195-4322
           USA 
           \and
           Donald~W.~Denbo \and Edison~Gica \and Harold~O.~Mofjeld \and Michael~C.~Spillane \and Vasily~V.~Titov \at
           NOAA/Pacific Marine Environmental Laboratory,
           7600 Sand Point Way NE,
           Seattle, WA  98115
           USA 
           \and
           Donald~W.~Denbo \and Edison~Gica \and Michael~C.~Spillane  \at
           Joint Institute for the Study of the Atmosphere and Ocean, University of Washington, Seattle, WA 98195-5672 USA
           \and
           Paul Y.~Huang \at
           National Tsunami Warning Center,
           National Weather Service,
           Palmer, AK, 99645
           USA 
}

\date{Received: date / Accepted: date}

\maketitle

\begin{abstract}

A performance measure for a \DART\ tsunami buoy network has been developed. The measure is based on a statistical analysis of simulated forecasts of wave heights outside an impact site and how much the forecasts are degraded in accuracy when one or more buoys are inoperative. The analysis uses simulated tsunami height time series collected at each buoy from selected source segments in the Short-term Inundation Forecast for Tsunamis (SIFT) database and involves a set for 1000 forecasts for each buoy/segment pair at sites just offshore of selected impact communities. Random error-producing scatter in the time series is induced by uncertainties in the source location, addition of real oceanic noise, and imperfect tidal removal. Comparison with an error-free standard leads to root-mean-square errors (RMSEs) for \DART\ buoys located near a subduction zone. The RMSEs indicate which buoy provides the best forecast (lowest RMSE) for sections of the zone, under a warning-time constraint for the forecasts of 3 hrs. The analysis also shows how the forecasts are degraded (larger minimum RMSE among the remaining buoys) when one or more buoys become inoperative. The RMSEs also provide a way to assess array augmentation or redesign such as moving buoys to more optimal locations. Examples are shown for buoys off the Aleutian Islands and off the West Coast of South America for impact sites at Hilo HI and along the U.S.~West Coast (Crescent City CA and Port San Luis CA).  A simple measure (coded green, yellow or red) of the current status of the network's ability to deliver accurate forecasts is proposed to flag the urgency of buoy repair.
 
\end{abstract}

\section{Introduction}
\label{intro}

Tsunamis are potentially devastating disasters for coastal regions worldwide.
The loss of life due to tsunamis can be greatly reduced
when the propagation time of a tsunami is long enough to allow for timely issuance of warnings.
Beginning soon after devastating tsunamis in 1960 and 1964, the National Oceanic and Atmospheric Administration (NOAA) and its predecessors have provided tsunami forecasts to the United States and other nations. The research presented here is part of an ongoing effort
to improve the speed and accuracy of such forecasts through enhanced observational capabilities.

Prompted in part by the scale of destruction and unprecedented loss of life
following the December 2004 Sumatra tsunami,
NOAA has deployed 39 Deep-ocean Assessment and Reporting of Tsunamis (\DART) buoys
(to date, eight countries other than the United States have set out additional buoys). 
These buoys can observe the passage of a tsunami
and relay data related to its arrival time and amplitude in near-real time.
These data are used by the Short-term Inundation Forecast for Tsunamis (SIFT) application
developed at the NOAA Center for Tsunami Research (NCTR);
for details, see Gica et al.~({\bl 2008}) and Titov ({\bl 2009}).
SIFT combines \DART\ data with precomputed geophysical models
to estimate so-called unit source coefficients (Percival et al.~{\bl 2011}).
As outlined in Sect.~{\bl\ref{sec:Construction}},
these coefficients are used to forecast wave heights at locations in the open ocean
outside of impact sites (coastal regions of particular interest, e.g., Hilo HI). 
In turn forecasts of these wave heights play a critical role in forecasting coastal inundation
at impact sites.

In this article we consider the problem of how to evaluate
the effectiveness of a \DART\ buoy network
in providing accurate and timely forecasts of open-ocean wave heights.
For an existing network
the goal is to devise a simple measure of effectiveness that can be monitored
over time and used to raise an alarm if the network deteriorates.
For an alteration to an existing network
(typically through adding, decommissioning or relocating a buoy),
the goal is to quantify the effect of the change in a simple manner.
The challenge in both cases is that
the effectiveness of wave height forecasts is predicated on a number of interacting factors.
The desire for a simple measure is at odds with the inherent complexity of the forecast process.
What is needed is a useful measure that does not oversimplify.
Here we propose a measure that makes certain simplifications
but nonetheless captures key factors limiting the accuracy of forecasted wave heights.
Our approach is to use a succession of summaries of simulated forecasts.
(An ideal measure would make use of actual rather than simulated forecasts,
but there is simply not enough actual data for realistic evaluation
of an existing network, much less a proposed network.)

Here is an outline of the rest of this article.
After a review in Sect.~{\bl\ref{sec:Construction}}
of how open-ocean wave height forecasts are constructed,
we discuss evaluating the quality of these forecasts
by (1)~simulating tsunami signals and corresponding open-ocean wave heights
(Sect.~{\bl\ref{subsec:SimTsunamiSignal}});
(2)~simulating the data collected by \DART\ buoys
(Sect.~{\bl\ref{subsec:SimOneMinStream}});
(3)~using the simulated data to estimate the source coefficients
that are needed to forecast open-ocean wave heights 
(Sect.~{\bl\ref{subsec:EstSourceCoeffs}}); and
(4)~using root-mean-square errors (RMSEs) as a criterion  
(Sect.~{\bl\ref{subsec:Evaluation}}).
In Sect.~{\bl\ref{sec:network}}
we illustrate use of these RMSEs for network evaluation through two case studies.
In the first study (Sect.~{\bl\ref{subsec:AleutianIslands}})
we focus on tsunami events arising from the Aleutian Islands
to evaluate how well the existing network of \DART\ buoys holds up
for forecasting wave heights outside of three impact sites
when one or more buoys become inoperative
(the sites are Hilo HI, Crescent City CA and Port San Luis CA).
We introduce a simple `green/yellow/red' indicator of the status
of the network (healthy/some deterioration/serious deterioration).
We use the second case study (Sect.~{\bl\ref{subsec:SouthAmerica}})
to demonstrate the use of our proposed methodology for assessing
the addition of a new buoy to an existing South American network.
The final two sections of the article contain some discussions (Sect.~{\bl\ref{sec:discussion}}),
a summary and our conclusions (Sect.~{\bl\ref{sec:conclusions}}).

\section{Construction of Open-Ocean Wave Height Forecasts in Practice}
\label{sec:Construction}

\begin{figure}
\centering
\includegraphics[width=4.5in,angle=0]{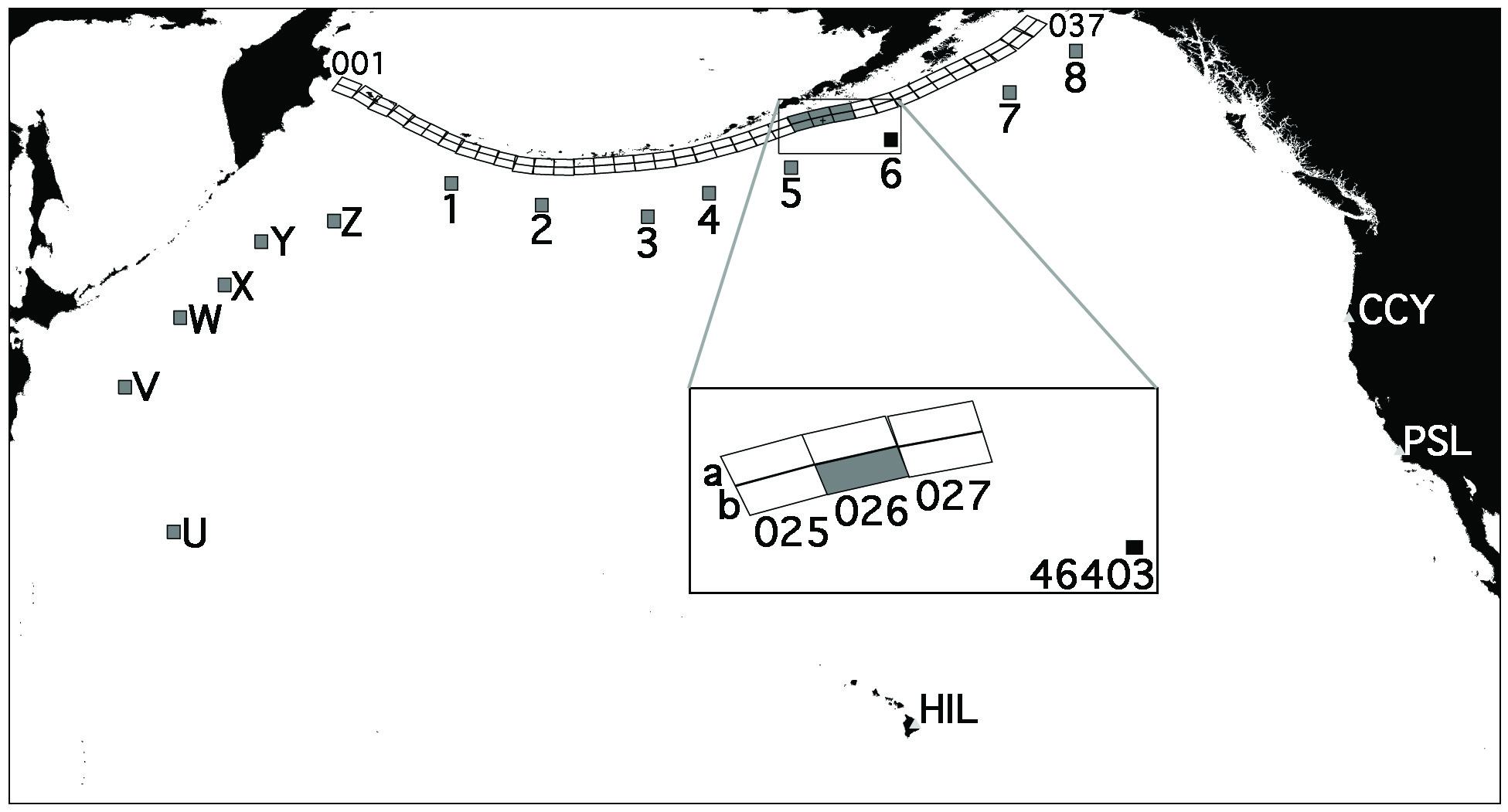}
\caption{
Locations of 74 unit sources in Aleutian--Alaskan subduction zone
(small rectangles, with an expanded view of six of them),
14 \DART\ buoys in the Northern Pacific (solid squares)
and three impact sites (HIL marks Hilo,
while CCY and PSL show where Crescent City and Port San Luis are).
The unit sources are arranged in two rows denoted by {\tt a} (upper row) and {\tt b} (lower)
and in 37 columns labelled as, from left to right, {\tt 001} to {\tt 037}.
Going clockwise, the World Meteorological Organization designations for the 14 buoys are
({\tt U})~21413,
({\tt V})~21418,
({\tt W})~21401,
({\tt X})~21419,
({\tt Y})~21402,
({\tt Z})~21416,
({\tt 1})~21415,
({\tt 2})~21414,
({\tt 3})~46413,
({\tt 4})~46408,
({\tt 5})~46402,
({\tt 6})~46403,
({\tt 7})~46409  and
({\tt 8})~46410.
The expanded view shows central unit source {\tt ac026b} and the five unit sources abutting it,
along with the location of the nearby buoy 46403
}
\label{fig:Aleutian}
\end{figure}

Here we give an overview of how open-ocean wave height forecasts are made in practice.
Generation of these forecasts presumes that
a tsunami-causing earthquake originates from within one or more unit sources,
which are 100 km by 50 km sections of fault planes in tsunamigenic regions.
Figure~{\bl\ref{fig:Aleutian}} shows
locations of a subset of 74 predesignated unit sources tiling the region
where the Pacific plate is being subducted
within the Aleutian--Alaska subduction zone.
The sources chosen are in 2 rows ({\tt b} and {\tt a}) and 37 columns ({\tt 001} to {\tt 037}).
Row {\tt b} is adjacent to the plate boundary,
while row {\tt a} is a continuation of the descending slab to greater depths
(we do not make use of additional rows used to represent deeper Aleutian--Alaska earthquakes).
The expanded view in the figure shows rows {\tt b} and {\tt a}
along with 3 of the columns ({\tt 025}, {\tt 026} and {\tt 027}).
We refer to the unit source in, e.g., row {\tt b} and column {\tt 026} as {\tt ac026b}. 
The figure also shows the locations of three impact sites
(Hilo (HIL), Crescent City (CCY) and Port San Luis (PSL))
and the current locations of 14 \DART\ buoys
(labeled clockwise by U, V, \dots, X, 1, 2, \dots, 8;
the expanded view traps buoy 46403, which is labeled as {\tt 6}).
There are 39 \DART\ buoys in all,
but these 14 are germane for handling events arising from Aleutian--Alaska earthquakes
in that, as called for by standard operating procedures,
their data could be used as the basis for issuing a warning at least three hours in advance
to at least one of the three impact sites
(Whitmore et al.~{\bl 2008}).

\begin{figure}
\centering
\includegraphics{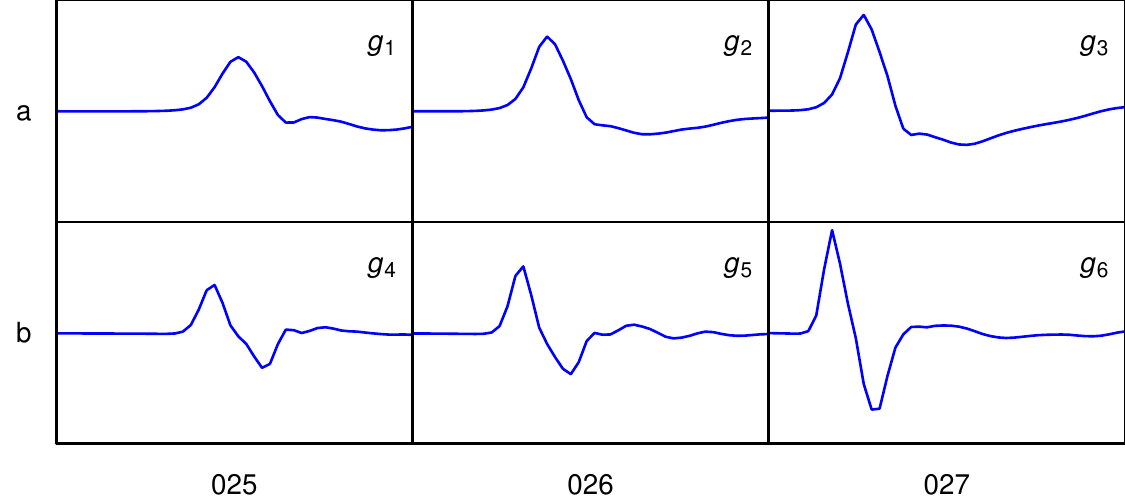}
\caption{Models for what would be observed at \DART\ buoy 46403
from an earthquake originating at the center of one of six unit sources
in the Aleutian--Alaskan subduction zone
(the 100 km by 50 km sections of the ocean corresponding to these unit sources
are highlighted in Fig.~{\bl\ref{fig:Aleutian}}).
The top row shows the models for unit sources
{\tt ac025a}, {\tt ac026a} and {\tt ac027a} (labeled as ${\bm g}_1$, ${\bm g}_2$ and ${\bm g}_3$);
the bottom,
for {\tt ac025b}, {\tt ac026b} and {\tt ac027b} (${\bm g}_4$, ${\bm g}_5$ and ${\bm g}_6$).
The time span for each displayed model is 45 minutes,
and the span starts 10 minutes after the start of the earthquake
}
\label{fig:SixUnitSources}
\end{figure}

A geophysical model is precomputed for each pairing
of one of the 74 unit sources with either a particular buoy
or a location in the open ocean outside of an impact site of interest
(these and other models are stored in a propagation database maintained at NCTR;
Gica et al.~{\bl 2008})
The model predicts what would be observed over time at a deep water location 
if a moment magnitude $M_W = 7.5$ reverse thrust fault earthquake
were to originate from within a particular unit source.
Figure~{\bl\ref{fig:SixUnitSources}} shows examples of predictions for buoy 46403
due to an earthquake from either unit source {\tt ac026b} or one of five abutting it
(for use later on, these predictions are marked by ${\bm g}_1$, \dots, ${\bm g}_6$).
If, as is usually the case,
the moment magnitude of the earthquake is different from 7.5,
then, relying on the linearity of tsunami waves in deep water,
the predictions are adjusted by a nonnegative multiplication factor $\alpha$
known as a source coefficient.
Earthquakes can span more than one unit source,
in which case the model is taken to be a linear combination
of individual unit source models,
with the coefficients in the linear combination,
say $\alpha_k$, $k=1,\ldots,K$, all constrained to be nonnegative.

During the passage of a tsunami,
a \DART\ buoy collects and transmits a time series of bottom pressure (BP) measurements.
Each measurement represents the BP averaged over a 1-min span.
The spans are non-overlapping, and the measurements occur once per minute,
thus forming what we refer to as a 1-min stream of data.
Once a sufficient stretch of this stream has been collected by the first buoy
that detects the tsunami event,
the source coefficients $\alpha_k$ are estimated using constrained least squares
(Percival et al.~{\bl 2014}).
With the passage of time,
more 1-min data from this and potentially other buoys become available,
and the estimates $\hat \alpha_k$ 
of the source coefficients can be reevaluated and possibly refined.
In operational tsunami forecasting,
the estimated source coefficients are applied to a sub-region of the propagation database
to provide the boundary forcing fields for the real-time inundation model run
with detailed simulation of waves, currents and inundation for the impact site
(Gica et al.~{\bl 2008});
however, in the simulation study discussed in the next section,
we avoid use of the inundation model
by choosing an adjacent deep-water location for each impact site of interest.
This approach allows us to efficiently generate a large numbers of simulations 
by linear combinations of table look-ups.

\section{Evaluation of Open-Ocean Wave Height Forecasts via Simulation}
\label{sec:Assessing}

In theory the best way to evaluate the effectiveness of a network of \DART\ buoys
would be to compare coastal inundation forecasts formulated
during actual tsunami events with the actual inundations.
While such comparisons are routinely done as a vital check on current operational procedures,
use of these alone is not sufficient for network evaluation.
A meaningful assessment would require not only multiple events,
but also ones treated in the same manner
(no change in the network of buoys, no equipment upgrades,
use of the same methodology for forming the forecasts, etc.).
Even if conditions for generating forecasts could be fixed,
the amount of time needed to collect data for a reliable evaluation
would be inordinate given the rate at which tsunamis occur.
In addition,
if we were to insist upon using actual events to assess effectiveness,
changes to a network could not be assessed
until long after the changes had been put into place.
For planning purposes,
we need the ability to assess the effectiveness
of any proposed changes.
We thus entertain basing our assessments on simulated forecasts.
 
As noted in the previous section,
the run-up model for generating inundation forecasts at a particular impact site
requires its initial conditions be set
by wave heights in the open ocean at a nearby location. 
The quality of the forecasts generated by the run-up model
depends critically on these initial conditions,
which are provided by forecasts of open-ocean wave heights.
Rather than assessing a network via the quality of inundation forecasts directly,
we can do so indirectly via the quality of open-ocean wave height forecasts.
In the simulation studies discussed below,
this indirect approach is computationally attractive
because calculation of run-up models is time-consuming.

With the simplification of dealing with open-ocean wave height forecasts
rather than inundation forecasts,
there are four tasks at hand -- discussed in the subsections to follow --
to simulate and evaluate the forecasts of interest.
The first task is to play the role of nature (Sect.~{\bl\ref{subsec:SimTsunamiSignal}}).
We must simulate a tsunami event along with what we will regard as the wave heights
presumed to occur outside of an impact site of interest.
The second task is to play the role of nature's observer (Sect.~{\bl\ref{subsec:SimOneMinStream}}).
We must simulate the data that \DART\ buoy would collect during a tsunami event.
The third task is to use the simulated data to estimate source coefficients,
which in turn are used to forecast the wave heights at an impact site
(Sect.~{\bl\ref{subsec:EstSourceCoeffs}}).
The fourth task is to evaluate the quality of the forecasted wave heights
by comparing them with the presumed heights using an appropriate metric
(Sect.~{\bl\ref{subsec:Evaluation}}).

In the remainder of this section,
we illustrate how we carry out these four tasks by focusing on a representative triad
consisting of one predesignated central unit source ({\tt ac026b}),
one \DART\ buoy (46403) and one impact site (Hilo HI).
The evaluation of the effectiveness of a \DART\ buoy network
requires a systematic look at other triads,
which is considered in Sect.~{\bl\ref{sec:network}}.

\subsection{Simulation of tsunami signal and open-ocean wave heights}
\label{subsec:SimTsunamiSignal}

\begin{figure}
\centering
\includegraphics{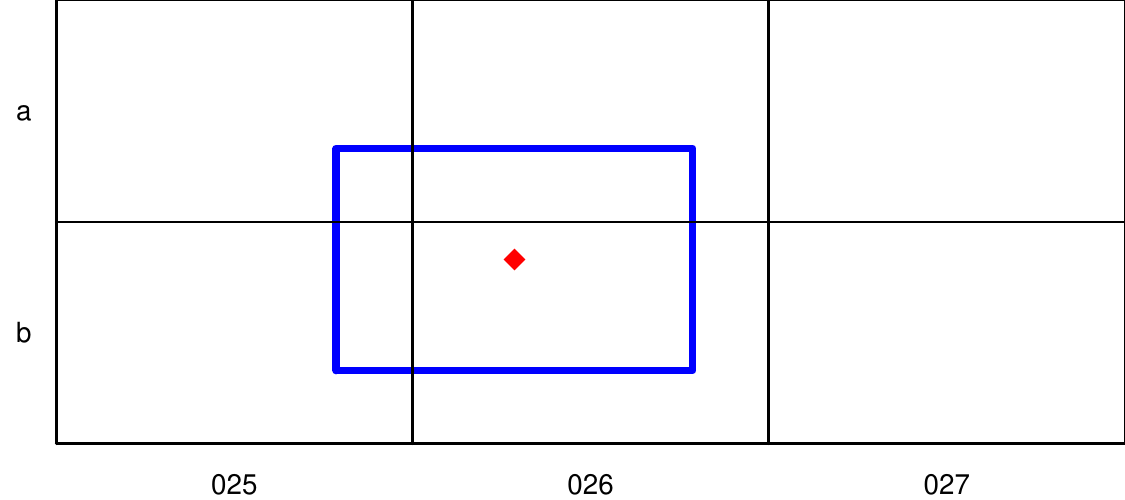}
\caption{Six rectangles (black lines) representing 100 km by 50 km sections of ocean
associated with predesignated unit sources
{\tt ac025a}, {\tt ac026a} and {\tt ac027a}
(top row)
and 
{\tt ac025b}, {\tt ac026b} and {\tt ac027b} (bottom)
in the Aleutian--Alaskan subduction zone
(these are highlighted in Fig.~{\bl\ref{fig:Aleutian}}).
The red diamond within the rectangle for the central unit source {\tt ac026b}
defines the center of a rectangle for a relocated unit source (blue lines).
The relocated rectangle intersects the rectangles
associated with {\tt ac025a}, {\tt ac026a}, {\tt ac025b} and {\tt ac026b}
(these are associated with models ${\bm g}_1$, ${\bm g}_2$, ${\bm g}_4$, ${\bm g}_5$
shown in Fig.~{\bl\ref{fig:SixUnitSources}})
}
\label{fig:RandomPick}
\end{figure}

Our first task is to simulate a tsunami event,
which manifests itself as a tsunami signal
that passes by a particular \DART\ buoy, say, 46403
(Fig.~{\bl\ref{fig:Aleutian}} shows the location of this buoy).
We denote this signal by ${\bm t}$.
A simple way to specify ${\bm t}$ would be to use
one of the precomputed models predicting
what 46403 should see if the tsunami event were to originate
from within a particular predesignated unit source, say, {\tt ac026b}
(shown in Fig.~{\bl\ref{fig:Aleutian}}).
We would thus just set ${\bm t}$ equal to ${\bm g}_5$
depicted in Fig.~{\bl\ref{fig:SixUnitSources}};
however, we have chosen {\it not\/} to adopt this simple way of simulating the signal
because an important factor
impacting open-ocean wave height forecasts is modeling of the tsunami signal ${\bm t}$
(this is the subject of Sect.~{\bl\ref{subsec:EstSourceCoeffs}}).
Because we cannot model the signal perfectly in an actual tsunami event,
there is a danger of generating unrealistic simulated forecasts
if we were to assume the signal to be identical to its model.
One source of mismatch is due to the uncertainty
in the location of the source within the subduction zone.
Figure~{\bl\ref{fig:RandomPick}} illustrates the effect of relocating unit source {\tt ac026b},
producing a relocated 100 km by 50 km source that overlaps (in this example)
with three additional predesignated sources.
To create the relocated source,
we picked a random point (the red diamond) within {\tt ac026b}
and used this point as the center for the relocated unit source (blue rectangle).
We can now compute a model predicting what would be observed at buoy 46403
if an earthquake were to originate from the randomly relocated unit source.
This new model will exhibit some degree of difference from the precomputed models
shown in Fig.~{\bl\ref{fig:SixUnitSources}}.
If we take this new model to be our tsunami signal ${\bm t}$
and if we were then to entertain modeling ${\bm t}$ using a subset of the models
shown in Fig.~{\bl\ref{fig:SixUnitSources}},
then the constructed signal need not be exactly equal to its model.
Randomly choosing a set of locations within a central unit source ({\tt ac026b} in this example)
then leads to an ensemble of relocated sources that reflects the uncertainty in source location.

\begin{figure}
\centering
\includegraphics{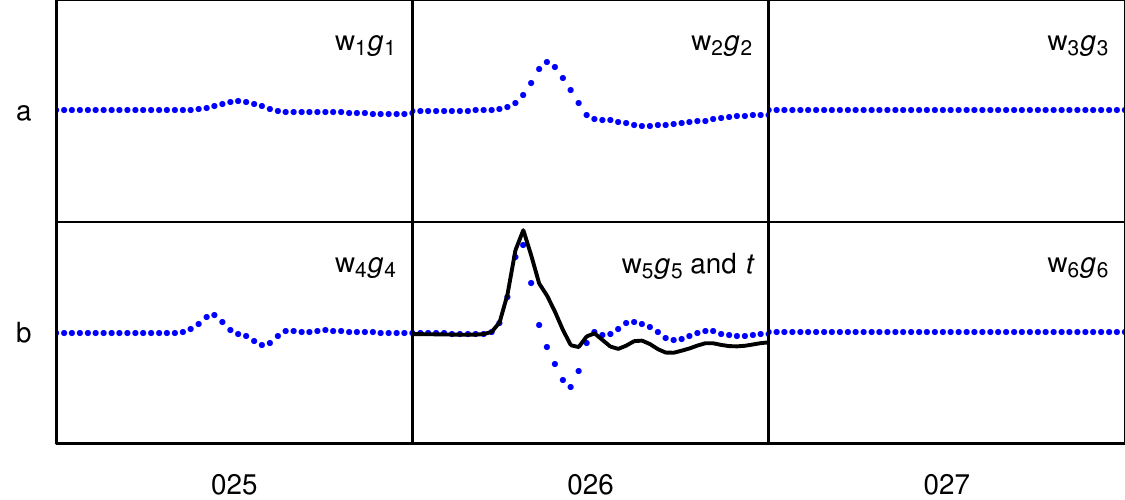}
\caption{Tsunami signal $\bm t$ (black curve, middle bottom plot)
formed from linear combination of models ${\bm g}_1, \ldots, {\bm g}_6$
as per Eq.~{\bl\ref{eq:SignalModel}} with
$w_1 \doteq 0.28$,
$w_2 \doteq 1.04$,
$w_3 = 0$,
$w_4 \doteq 0.57$,
$w_5 \doteq 2.11$
and
$w_6 = 0$
(the ${\bm g}_k$ models are shown in Fig.~{\bl\ref{fig:SixUnitSources}}).
The blue dotted curves show $w_k {\bm g}_k$,
and the sum of all six of these is equal to $\bm t$
(because $w_5=w_6=0$,
models ${\bm g}_5$ and ${\bm g}_6$ do not contribute to $\bm t$)}
\label{fig:RelocatedUnitSource}
\end{figure}

Because it is time consuming to generate geophysical models,
we make an assumption of linearity and construct the tsunami signal ${\bm t}$
using a linear combination of precomputed models.
For the example shown in Fig.~{\bl\ref{fig:RandomPick}},
the signal would be a linear combination of the models
for unit sources {\tt ac025a}, {\tt ac026a}, {\tt ac025b} and
{\tt ac026b} (the central unit source).
These models are labeled in Fig.~{\bl\ref{fig:SixUnitSources}}
as ${\bm g}_1$, ${\bm g}_2$, ${\bm g}_4$ and ${\bm g}_5$.
The weights $w_k$ in the linear combination are dictated by the degree of intersection
of the relocated source with the predesignated sources --
in this example, ${\bm g}_5$ (corresponding to {\tt ac026b}) gets the most weight,
while ${\bm g}_1$ (corresponding to {\tt ac025a}) gets the least.
Figure~{\bl\ref{fig:RelocatedUnitSource}} illustrates
the construction of the $N$-dimensional column vector ${\bm t}$
in terms of the six vectors ${\bm g}_k$ shown in Fig.~{\bl\ref{fig:SixUnitSources}}:
\begin{equation}\label{eq:SignalModel}
{\bm t} = \sum_{k=1}^6 w_k {\bm g}_k
\end{equation}
(black curve, lower middle plot).
Note that, since neither ${\bm g}_3$ nor ${\bm g}_6$ are involved in constructing
${\bm t}$, the weights $w_3$ and $w_6$ are zero.
Past experience with actual tsunami events suggests
that the weights should not be arbitrarily set
if we want to regard the simulated event
as originating from a 100~km by 50~km region
(Hanks and Kanamori {\bl 1979}; Papazachos et al.~{\bl 2004}).
We use the normalization $\sum_k w_k = 4$ in all of our simulations
because a larger setting than this would result in a tsunami signal
whose magnitude would be more realistically associated with multiple unit sources.
For the example shown in Fig.~{\bl\ref{fig:RandomPick}},
we have $w_1 \doteq 0.28$, $w_2 \doteq 1.04$, $w_4 \doteq 0.57$ and $w_5 \doteq 2.11$. 

\begin{figure}
\centering
\includegraphics{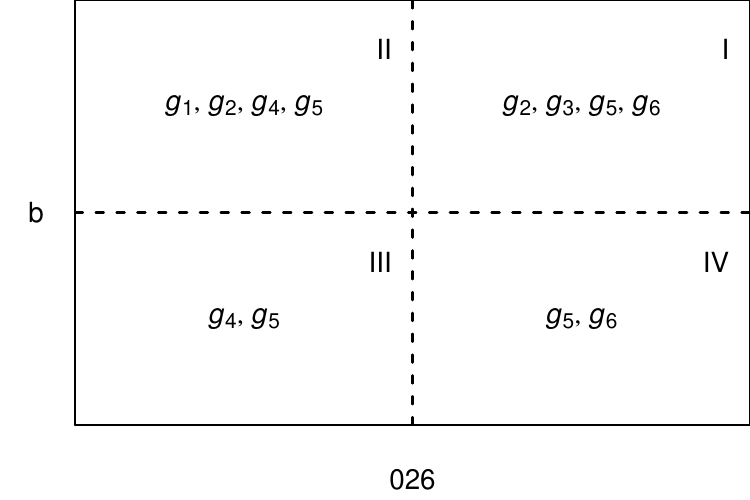}
\caption{Division of rectangle for central unit source {\tt ac026b} into four quadrants
(I, II, III and IV).
A point chosen at random within this rectangle is equally likely to fall
in each of the four quadrants.
The models ${\bm g}_k$ used to construct the tsunami signal $\bm t$
depend upon which quadrant the random pick falls in --
these models are listed in the middles of the quadrants.
The model ${\bm g}_5$ for {\tt ac026b} is always part of the constructed tsunami signal.
The weight $w_k$ assigned to each model depends upon the location of the random pick
within the quadrant, with the weight for ${\bm g}_5$ increasing as the random pick
gets closer to the center of the rectangle
}
\label{fig:FourQuads}
\end{figure}

We can construct additional tsunami signals
by selecting other points at random within the rectangle for the central unit source {\tt ac026b}.
Figure~{\bl\ref{fig:FourQuads}} breaks the rectangle for {\tt ac026b} up
into four quadrants of equal size (labeled as I, II, III and IV).
The specific models ${\bm g}_k$ that are used to construct the tsunami signal
depend upon which quadrant the randomly selected point falls in
-- these are indicated in Fig.~{\bl\ref{fig:FourQuads}} in the middle of each quadrant.
The random pick in Fig.~{\bl\ref{fig:RandomPick}} falls in quadrant II,
and hence, as previously noted,
the signal is constructed using ${\bm g}_1$, ${\bm g}_2$, ${\bm g}_4$ and ${\bm g}_5$.
No matter in which quadrant the random pick falls,
the model ${\bm g}_5$ is always included with a weight $w_5$ satisfying $1 \le w_5 \le 4$,
and this weight is close to 4 when the random pick is close to the center of the rectangle.
The upper two quadrants involve three models in addition to ${\bm g}_5$,
whereas the lower quadrants involve just one (either ${\bm g}_4$ or ${\bm g}_6$).
The characteristics of the constructed tsunami signals
will thus depend upon the randomly selected quadrant
and where the randomly selected point occurs within the quadrant.
No matter where the random pick ends up,
we can write the resulting constructed tsunami signal as per Eq.~({\bl\ref{eq:SignalModel}}),
with the understanding that either two or four of the $w_k$'s will be zero.

We can also construct tsunami signals by focusing on a unit source other than {\tt ac026b};
however,
for simplicity, we will use as a central unit source
only those having a setup similar to {\tt ac026b},
namely, that they are in row b and are surrounded by five other unit sources.
This restriction for central unit sources simplifies computer code somewhat
and should not adversely impact the evaluation of the effectiveness of a buoy network.

Given a relocated unit source,
there are two methods for generating the corresponding open-ocean wave heights ${\bm h}$
outside of an impact site of interest.
The more accurate method is via high-resolution model runs, which are time consuming;
the less accurate makes use of lower-resolution runs
already available in a precomputed propagation data base,
which has the advantage of being easy to extract.
We tested both methods and found that results based on the propagation data base
differed little from those based on high-resolution model runs.
We have thus elected to use the propagation data base approach,
for which we take the presumed wave heights to be
\begin{equation}\label{eq:ActualWaveHeights}
{\bm h} = \sum_{k=1}^6 w_k {\bm h}_k,
\end{equation}
where the weights $w_k$ are identical to those in Eq.~{\bl\ref{eq:SignalModel}},
while ${\bm h}_k$ is the forecast of the wave heights in the open ocean
that would occur from an event occurring in the same predesignated unit source
associated with ${\bm g}_k$
(note, however, that, while ${\bm h}_k$ depends upon the location of the unit source
and the location in the open ocean,
it does not depend upon {\it any\/} of the buoy locations).

\begin{figure}
\centering
\includegraphics{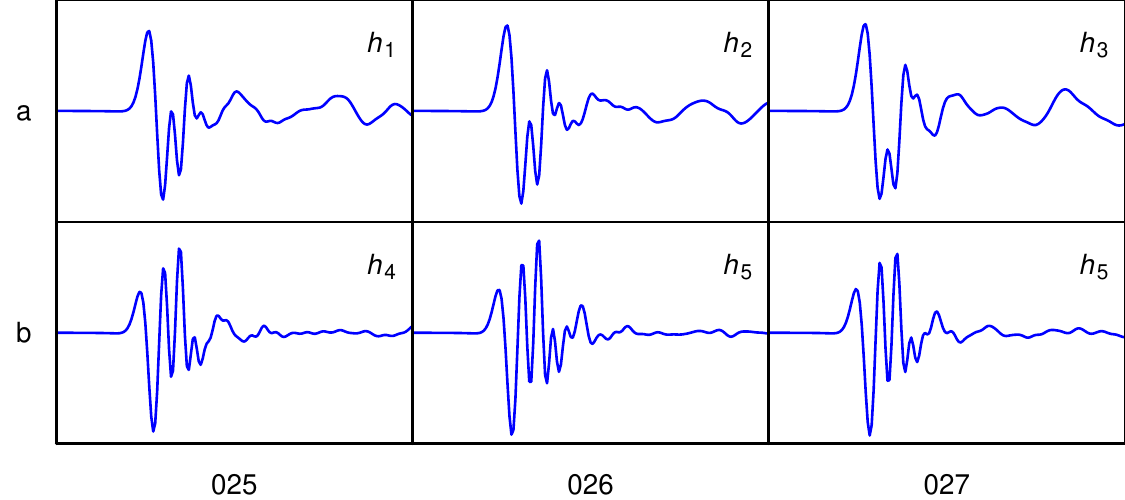}
\caption{Models ${\bm h}_1$, \dots, ${\bm h}_6$ for the wave heights
in the open ocean outside of Hilo from an earthquake
originating from either the predesignated central unit source {\tt ac026b}
(corresponding to ${\bm h}_5$)
or one of five adjacent unit sources
{\tt ac025a} (corresponding to ${\bm h}_1$),
{\tt ac026a} (${\bm h}_2$),
{\tt ac027a} (${\bm h}_3$),
{\tt ac025b} (${\bm h}_4$)
or
{\tt ac027b} (${\bm h}_6$)
located in the Aleutian--Alaskan subduction zone
(Fig.~{\bl\ref{fig:Aleutian}} shows the locations of the 100 km by 50 km sections
of the ocean associated with these unit sources).
The time span for each displayed model is three hours,
and the span starts at four hours from the beginning of the earthquake}
\label{fig:SixHiloWaveHeights}
\end{figure}

\begin{figure}
\centering
\includegraphics{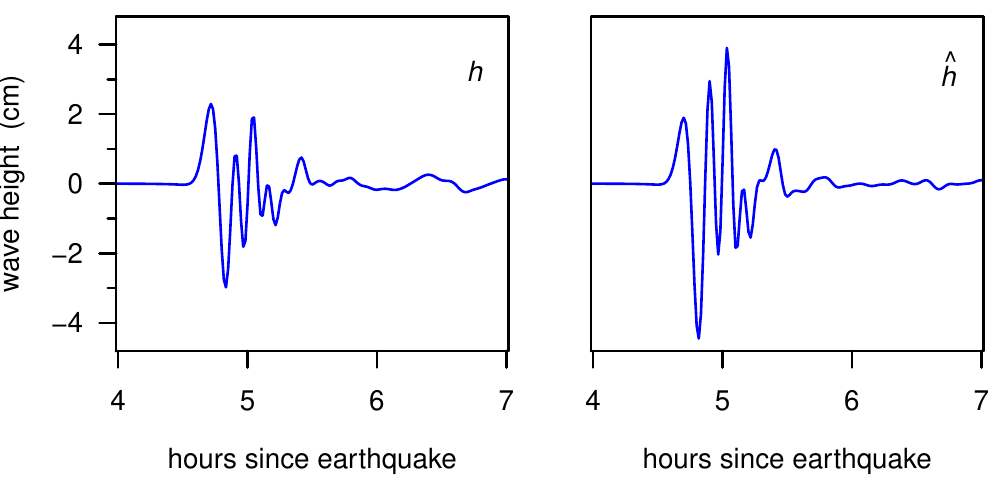}
\caption{Open-ocean wave heights ${\bm h}$ outside of Hilo
as dictated by relocated unit source of Fig.~{\bl\ref{fig:RandomPick}}
(left-hand plot)
and associated forecasted wave heights $\hat {\bm h}$
(right-hand).
The heights ${\bm h}$ are given as per Eq.~{\bl\ref{eq:ActualWaveHeights}}
with the same setting for $w_k$ as listed in the caption
to Fig.~{\bl\ref{fig:RelocatedUnitSource}}.
The heights $\hat {\bm h}$ are given as per Eq.~{\bl\ref{eq:ForecastedWaveHeights}}
with ${\cal K} = \{ 4, 5\}$,
$\hat \alpha_4 \doteq 1.33$ and $\hat \alpha_5 \doteq 2.10$
}
\label{fig:TwoHiloWaveHeights}
\end{figure}

As an example,
Fig.~{\bl\ref{fig:SixHiloWaveHeights}} shows wave heights ${\bm h}_k$
at an open-ocean location outside of Hilo associated with
the same six predesignated unit sources considered in
Figs.~{\bl\ref{fig:Aleutian}} and~{\bl\ref{fig:SixUnitSources}}.
The left-hand plot of Fig.~{\bl\ref{fig:TwoHiloWaveHeights}} shows ${\bm h}$
formed using the weights stated in the caption to Fig.~{\bl\ref{fig:RelocatedUnitSource}}
(the right-hand plot is explained in Sect.~{\bl\ref{subsec:EstSourceCoeffs}}).

\subsection{Simulation of 1-min stream}
\label{subsec:SimOneMinStream}
Our second task is to simulate data as it would be recorded at a \DART\ buoy.
Due both to the complexity of recorded tsunami data
and in the way in which these data are used to create wave height forecasts, 
simulations of forecasts that are fully accurate
are difficult and time consuming to create.
The approach we take is to impose simplifications
that nonetheless retain the salient factors
limiting the ability of a buoy network to generate accurate forecasts.
As a starting point,
we assume the 1-min stream recorded at a particular buoy is given by
\begin{equation}\label{eq:BasicModel}
\bar {\bm y} = {\bm x} + {\bm t} + {\bm \epsilon},
\end{equation}
where $\bar {\bm y}$ is a vector containing $N$ consecutive values from the 1-min stream;
${\bm x}$ represents tidal fluctuations;
$\bm t$ is the tsunami signal constructed as per Eq.~({\bl\ref{eq:SignalModel}});
and $\bm \epsilon$ is background noise.
Tidal fluctuations and background noise in the \DART\ data
are two factors that can adversely impact inundation forecasts,
so it is important to handle these realistically.
Rather than simulating these factors,
we can use archived \DART\ data
that were recorded under ambient conditions
and retrieved during routine servicing of the buoy.
The tsunami signal in Eq.~({\bl\ref{eq:BasicModel}}) is missing during ambient conditions,
so random samples from historical data can serve to generate $\bm x + \bm \epsilon$
(for details about the sampling procedure, see Sect.~3, Percival et al.~{\bl 2015}).
A complication is that not all currently deployed buoys have associated archived data
(and proposed buoys certainly don't). 
To handle such buoys, we use data from a surrogate buoy with good matching
oceanographic conditions.

\begin{figure}
\centering
\includegraphics{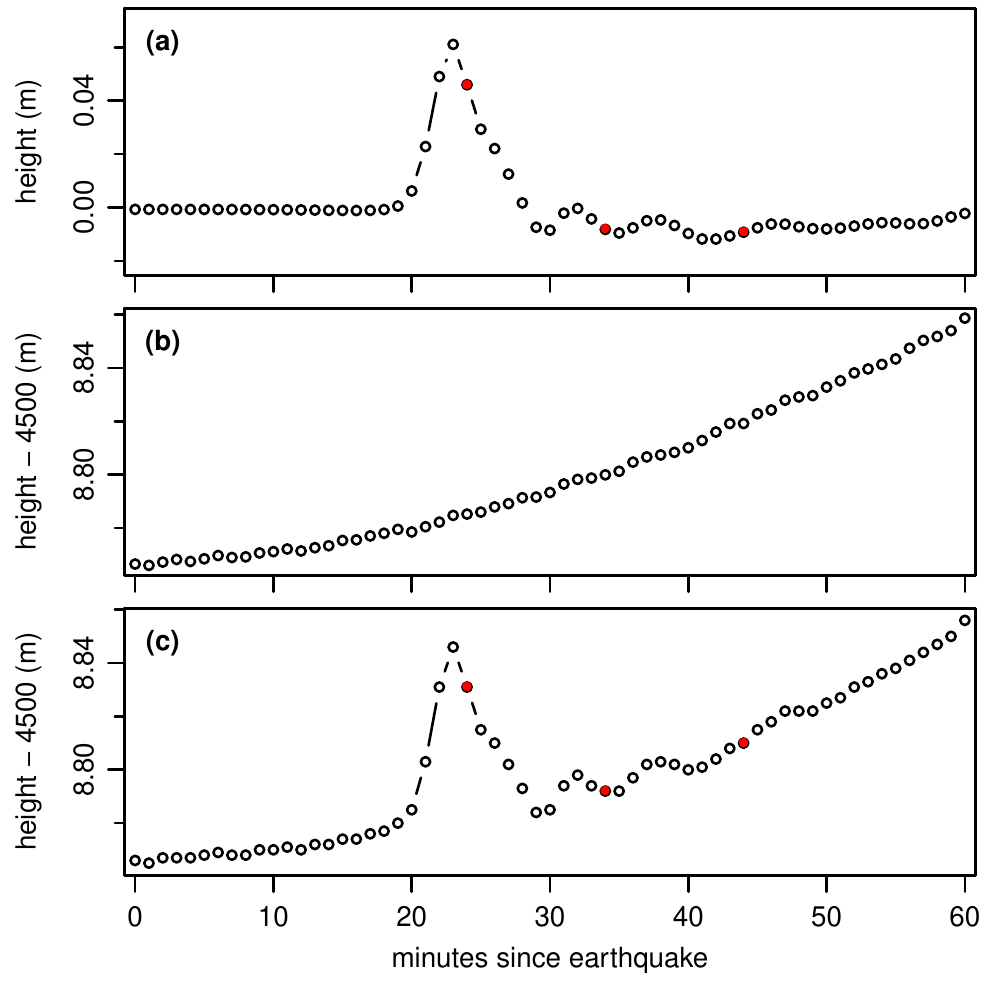}
\caption{Construction of a simulated 1-min stream $\bar {\bm y}$.
Plot~{\bf (a)} is the simulated tsunami signal --
this is the same as black curve in middle bottom plot of Fig.~{\bl\ref{fig:RelocatedUnitSource}}.
The three red circles indicate data occurring
1, 11 and 21 minutes after the signal's peak value.
Plot~{\bf (b)} shows tidal fluctuations and background noise, i.e., ${\bm x} + {\bm \epsilon}$,
recorded by \DART\ buoy 46403 under ambient conditions.
Plot~{\bf (c)} is $\bar {\bm y}$,
which is the sum of the time series in plots~{\bf (b)} and~{\bf (c)}
}
\label{fig:figOneSimulation}
\end{figure}

Figure~{\bl\ref{fig:figOneSimulation}} shows a simulated 1-min stream $\bar {\bm y}$
constructed as per Eqs.~({\bl\ref{eq:SignalModel}}) and~({\bl\ref{eq:BasicModel}}).
The top plot shows the tsunami signal $\bm t$
(this is the same as the black curve
in middle bottom plot of Fig.~{\bl\ref{fig:RelocatedUnitSource}}).
The middle plot shows the sum of tidal fluctuations $\bm x$ and background noise $\bm \epsilon$,
which were obtained from archived data for \DART\ buoy 46403.
The bottom plot shows $\bar {\bm y}$,
which is the sum of the time series in the top and middle plots.

\subsection{Estimation of source coefficients and forecasting of open-ocean wave heights}
\label{subsec:EstSourceCoeffs}
With simulated 1-min stream $\bar {\bm y}$ in hand, 
the third task is to estimate the source coefficients $\alpha_k$.
Following what is currently done within SIFT,
we assume a linear regression model of the form
\begin{equation}\label{eq:RegressionModel}
\bar {\bm y} =
\mu {\bm 1} + \beta_1 {\bm c} + \beta_2 {\bm s} + \sum_{k\in{\cal K}} \alpha_k {\bm g}_k + {\bm e},
\end{equation}
where $\bm 1$ is an $N$-dimensional vector of ones
associated with the regression coefficient $\mu$;
${\bm c}$ is a vector with elements $\cos\,(\omega n\,\Delta)$,
$n=0,1, \ldots, N-1$, and is associated with $\beta_1$
-- here $\omega$ is the tidal frequency M2 and $\Delta = 1$~min;
${\bm s}$ is associated with $\beta_2$ and is similar to ${\bm c}$
but its elements are given by $\sin\,(\omega n\,\Delta)$;
$\cal K$ is a subset of $\{ 1, 2, \ldots, 6 \}$
that specifies the ${\bm g}_k$ to be used to model the tsunami signal;
and ${\bm e}$ is a vector of stochastic errors assumed to have zero mean
(note that this vector is {\it not\/}
taken to be the same as ${\bm \epsilon}$ in Eq.~({\bl\ref{eq:BasicModel}})).
As discussed in Percival et al.~({\bl 2015}),
the coefficients $\mu$, $\beta_1$ and $\beta_2$ and their associated vectors
serve to model the tidal fluctuations in Eq.~({\bl\ref{eq:BasicModel}})
in a simple -- but statistically efficient -- manner.
Modeling of the signal $\bm t$ is handled through specification of $\cal K$.
We consider four protocols for setting $\cal K$.
\begin{enumerate}
\item[a.]
The {\it single unit source protocol\/} (`1 protocol' for brevity) sets ${\cal K}$ to be $\{ 5 \}$;
i.e., the model for the signal is $\alpha_5 {\bm g}_5$.
With probability one, this protocol yields an incorrect signal model
in the sense that the constructed signal consists of ${\bm g}_5$
combined with either one or three additional ${\bm g}_k$'s,
whereas our model for it involves just ${\bm g}_5$.
This protocol in effect assumes we only know
within which predesignated unit source
the earthquake occurred (i.e., we know nothing about the random pick).
\item[b.]
The {\it two unit sources protocol\/} (`2 protocol') assumes
we have limited information about the random pick,
namely, whether the earthquake originates from the left-hand or right-hand side of the rectangle
(the random pick in Fig.~~{\bl\ref{fig:RandomPick}} is left-handed).
For a left-hand pick, we set ${\cal K}$ to $\{ 4, 5 \}$,
and we set it to $\{ 5, 6 \}$ for a right-hand pick. 
For both settings, there are two ${\bm g}_k$'s in the model.
With this protocol,
there is a 50\% chance of having a mismatch
between the constructed signal and its model,
and the nature of the mismatch is that the model has two too few unit sources.
\item[c.]
The {\it four unit sources protocol\/} `4 protocol') also assumes
we know the left- or right-hand location of the earthquake.
Now we set ${\cal K}$ to $\{ 1, 2, 4, 5 \}$ for a left-hand pick and to $\{ 2, 3, 5, 6 \}$ otherwise.
There is again a 50\% chance of having a mismatch,
but now the nature of the mismatch is that there are two too many unit sources.
\item[d.]
The {\it matched protocol\/} sets ${\cal K}$
so that it contains {\it exactly\/} the indices used in forming the constructed signal;
i.e., we use the same set of ${\bm g}_k$'s both to form the signal and to model it
so there is no mismatch.
This protocol essentially presumes knowledge of the quadrant in which the random pick falls.
\end{enumerate}

After selection of the protocol ${\cal K}$,
the regression model of Eq.~({\bl\ref{eq:BasicModel}}) is now fully specified.
The model involves three coefficients for the tidal model ($\mu$, $\beta_1$ and $\beta_2$)
and either one, two or four source coefficients $\alpha_k$ for the signal model.
We estimate the regression coefficients using constrained least squares;
i.e., the estimated coefficients $\hat \mu$, $\hat \beta_1$, $\hat \beta_2$ and $\hat \alpha_k$
are those minimizing
\[
\big\| \bar {\bm y} - \mu {\bm 1} - \beta_1 {\bm c} - \beta_2 {\bm s} - \sum_{k\in{\cal K}} \alpha_k {\bm g}_k \big\|_2^2
\enskip\hbox{subject to $\alpha_k \ge 0$ for all $k$},
\]
where $\| {\bm x} \|_2^2 = \sum_n x^2_n$
is the squared Euclidean norm of a vector ${\bm x}$ with elements $x_n$. 
The non-negativity constraint on each source coefficient is critical
because, without it,
there is nothing to prevent the least square estimate of $\alpha_k$ from being negative,
which would render it inconsistent with the type of generating event
we are assuming (a reverse thrust earthquake).

Another important factor that impacts the quality of the inundation forecasts
is the amount of data $N$ available for estimating the source coefficients.
In general, more data in $\bar {\bm y}$ improves the quality of the estimated coefficients,
which, in turn, improves inundation forecasts.
To study the effect of $N$,
we consider three settings intended to loosely mimic
what would be available during different stages of an ongoing tsunami event.
These settings are based on the location of the first peak of the constructed signal $\bm t$
(alternatively we could use the first peak of the noisy data $\bar {\bm y}$,
but, to avoid issues that arise in automatic detection of a peak possibly distorted by noise,
we have chosen to use the constructed signal instead). 
The first setting is one minute past the first peak in the signal,
and the second and third settings are 11 and 21 minutes past the peak.
The three settings are illustrated in Figs.~{\bl\ref{fig:figOneSimulation}(a)} and~(c)
using red circles.
For this particular example,
the settings correspond to 24, 34 and 44 minutes after the start of the earthquake,
and we would estimate the coefficients based on placing
the corresponding first $N=25$, 35 or 45 simulated buoy measurements in~(c)
into $\bar {\bm y}$.
For tsunami signals other than this example,
we identify the first peak and the same related three points,
but only use a maximum of 60 minutes worth of data prior to the first peak
in cases where this peak occurs more than an hour after the start of the earthquake
(this mimics the amount of the 1-min stream available during an actual event). 

As an example of how the sample size influences source coefficient estimation,
let us focus on the simulated \DART\ buoy data $\bar {\bm y}$
shown in Fig.~{\bl\ref{fig:figOneSimulation}(c)}
and set $\cal K$ as per the 2 protocol.
Because the random pick is left-handed,
the model for the tsunami signal becomes $\alpha_4 {\bm g}_4 + \alpha_5 {\bm g}_5$.
There is thus a mismatch between the model and signal
since the latter makes use of ${\bm g}_1$ and ${\bm g}_2$ in addition to ${\bm g}_4$ and ${\bm g}_4$.
The following table shows the estimated source coefficients
for the three sample sizes along with the presumed weights $w_4$ and $w_5$.
\[
\hskip102pt
\vbox{
\halign{
\hfil#\hfil\hskip5.75pt&
\hfil#\hfil\hskip7pt&
\hfil#\hfil\hskip4.75pt&
\hfil#\hfil\hskip4.75pt&
\hfil#\hfil\hskip4.75pt&
\hfil#\hfil\hskip4.75pt&
\hfil#\hfil\cr
\noalign{\hrule\smallskip\hrule\medskip}
\noalign{\hskip35.25pt \hbox to 271pt{\hskip39pt$\hat \alpha_k$\hfill}}
\noalign{
\medskip
\nointerlineskip\moveright 39.00pt\vbox{\hrule width 104.25pt}\nointerlineskip
\smallskip}
$k$&$w_k$&&$N=25$&$N=35$&$N=45$\cr
\noalign{\smallskip\hrule\smallskip}
  4&0.57&&88.92&1.63&1.33\cr
  5&2.11&& 2.51&2.34&2.10\cr
\noalign{\smallskip\hrule\smallskip\hrule}
}
}
\]
The estimates improve with increasing $N$ in the sense that
$\hat \alpha$ gets closer to $w_k$ for both $k=4$ and $5$.

The forecasted wave heights are
\begin{equation}\label{eq:ForecastedWaveHeights}
\hat {\bm h} = \sum_{k\in{\cal K}} \hat \alpha_k {\bm h}_k.
\end{equation}
For the 2 protocol with a left-handed pick,
these heights are given by
\[
\hat {\bm h} = \hat \alpha_4 {\bm h}_4 + \hat \alpha_5 {\bm h}_5,
\]
where ${\bm h}_4$ and  ${\bm h}_5$
are depicted in Fig.~{\bl\ref{fig:SixHiloWaveHeights}}.
Use of the estimates for $\alpha_k$ corresponding to $N=45$
in the table above leads to the forecasted wave heights
shown in the right-hand plot of Fig.~{\bl\ref{fig:TwoHiloWaveHeights}}. 

To summarize,
estimates of the source coefficients $\alpha_k$ are imperfect in practice
due to major factors including, inter alia,
\begin{enumerate}
\item an imperfect tidal model;
\item background noise;
\item a mismatch between the assumed model and the actual tsunami signal;
\item a limited amount of data and
\item seismic noise, which is potentially important, but, because of timing,
often does not come into play. 
\end{enumerate}
Our simulation study takes into account 1--4, but ignores seismic noise.

\subsection{Evaluation of forecasted open-ocean wave heights}
\label{subsec:Evaluation}
Our final task is to quantify 
how well the forecasted wave heights $\hat {\bm h}$ match the presumed heights ${\bm h}$.
We considered three metrics.
The first is the maximum cross-correlation
between $\hat {\bm h}$ and either ${\bm h}$ or a lagged versions thereof;
the second is the squared difference between the maximum heights
in $\hat {\bm h}$ and ${\bm h}$;
and the third is the squared difference
between the maximum four-hour `energies' in $\hat {\bm h}$ and ${\bm h}$,
where the energies in question are the sum of squares of data
over all possible four-hour stretches contained within $\hat {\bm h}$ or ${\bm h}$.
The latter two metrics are of more operational interest than maximum cross-correlation,
which is also the most sensitive of the three to
innocuous misalignments in time between $\hat {\bm h}$ and ${\bm h}$.
In tests to date,
we have found that 
use of either maximum wave heights or maximum energies leads
to evaluations of network effectiveness that are qualitatively similar.
Since maximum wave heights are less time consuming to compute,
we stick with these in all that follows.
For the example shown in Fig.~{\bl\ref{fig:TwoHiloWaveHeights}},
we have $\max\,\{ {\bm h} \} \doteq 2.3$~cm and $\max\,\{ \hat {\bm h} \} \doteq 3.9$~cm,
and the metric is $(\max\,\{\hat {\bm h} \} - \max\,\{{\bm h} \})^2 \doteq 2.6$~$\hbox{cm}^2$.

A more thorough assessment of how well we can forecast wave heights outside of Hilo
requires repeating what is set forth in the previous three subsections over and over again.
We do so by generating 1000 different tsunami signals $\bm t$ as per Eq.~{\bl\ref{eq:SignalModel}}.
The signals are associated with 1000 independent random picks
from within unit source {\tt ac026b}.
Each pick serves as a center for a relocated unit source
whose location dictates the weights $w_k$ used to form both $\bm t$
and the presumed open-ocean wave heights $\bm h$ of Eq.~{\bl\ref{eq:ActualWaveHeights}}.
Each signal $\bm t$ is then added to an independently drawn random sample
of data recorded under ambient conditions by an appropriate \DART\ buoy. 
This addition forms the simulated 1-min stream $\bar{\bm y}$ as per Eq.~{\bl\ref{eq:BasicModel}}.
For each realization,
we then compute twelve sets of estimated source coefficients $\hat \alpha_k$
corresponding to twelve formulations of the linear model of Eq.~{\bl\ref{eq:RegressionModel}}.
The formulations involve the four protocols for setting $\cal K$
in combination with the three ways of determining the amount of data $N$ to be placed in $\bar {\bm y}$.
For a given formulation, we use the source coefficient estimates $\hat \alpha_k$
to generate -- as per Eq.~{\bl\ref{eq:ForecastedWaveHeights}} -- forecasted wave heights $\hat {\bm h}$
outside of the chosen impact site (Hilo in our representative triad).
We then compare the peak values in ${\bm h}$ and $\hat {\bm h}$.
\begin{figure}
\centering
\includegraphics{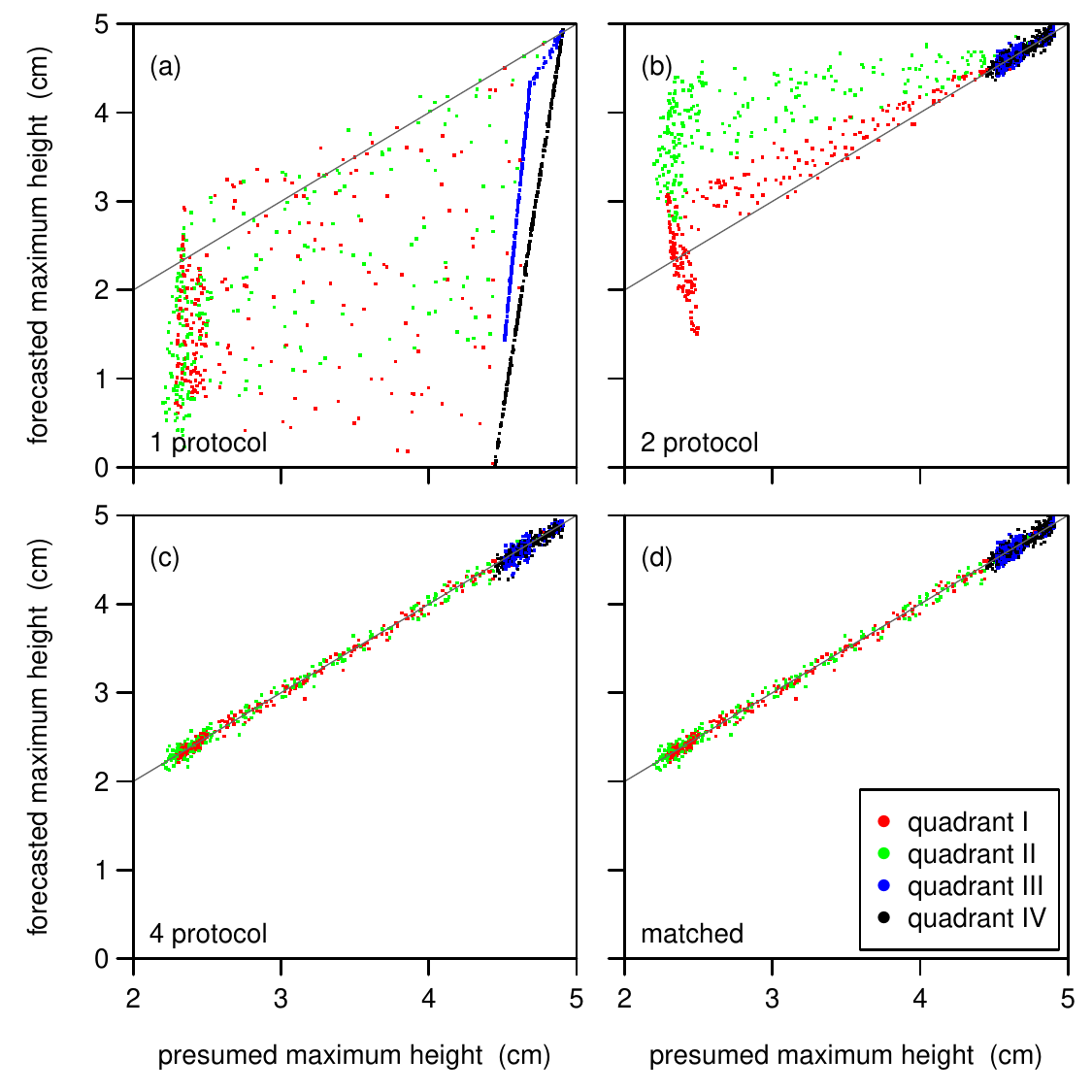}
\caption{Scatterplots of forecasted versus presumed maximum wave heights
for 1000 realizations of representative triad
(central unit source {\tt ac026b}, buoy 46403 and impact site Hilo).
Plots (a), (b), (c) and (d) correspond to the 1, 2, 4 and matched protocols
for fully specifying the regression model of Eq.~({\bl\ref{eq:RegressionModel}}).
Each realization is based on a random pick 
that has an equal chance of falling within one of the four quadrants
shown in Fig.~{\bl\ref{fig:FourQuads}},
with the quadrant determining which
models ${\bm g}_k$ receive nonzero weights $w_k$
in constructing the signal $\bm t$ via Eq.~({\bl\ref{eq:SignalModel}}).
As per the legend on plot~(d),
the colors of the points in the scatterplots
indicate into which quadrant the random pick fell.
The diagonal line in each plot shows where a point would fall
if the forecasted and presumed heights were in perfect agreement.
The amount of data $N$ used in the regression model is dictated by 21 minutes
past the first peak in the constructed signal $\bm t$
}
\label{fig:ScatterPlots}
\end{figure}

Figure~{\bl\ref{fig:ScatterPlots}} shows
forecasted maximum heights versus corresponding presumed heights
for 1000 realizations of our representative triad
(central unit source {\tt ac026b},
\DART\ buoy 46403
and a location in the open ocean outside of the impact site Hilo).
Each realization corresponds to a single point in each of the four plots.
In addition to the 1000 points,
each plot has a diagonal line indicating
where a point would fall if the forecasted and presumed heights were in perfect agreement.
For all four plots,
we set $N$ such that 21 minutes worth of data past the peak value in the signal $\bm t$
is used to estimate the source coefficients
and other parameters in the regression model (Eq.~({\bl\ref{eq:RegressionModel}})).
Plots~(a) to~(d) correspond to, respectively,
the 1, 2, 4 and matched protocols $\cal K$ for fully specifying the model.

The colors of the points in Fig.~{\bl\ref{fig:ScatterPlots}} indicate the quadrant
in which the random pick for a particular realization fell.
Of the 1000 picks,
245 were in quadrant I (colored red);
247, in II (green);
268, in III (blue);
and 240, in IV (black).
This distribution is not inconsistent
with each pick being equally likely to fall in any of the four quadrants.
As noted in Fig.~{\bl\ref{fig:FourQuads}},
picks in quadrants I and II involve linear combinations of four ${\bm g}_k$'s.
For the 1 and 2 protocols,
forecasted heights are thus based on mismatched models.
The mismatch is due to relevant ${\bm g}_k$'s being left out of the regression model.
As a result, the red and green points in plots~(a) and~(b)
have prominent scatter off the diagonal line. 
Plots~(c) and~(d) indicate that the forecasts are markedly better
when there is no model mismatch,
as occurs in the 4 and matched protocols
(note that the red and green points
in~(c) and~(d) are necessarily identical). 
In the absence of model mismatch, 
the remaining inaccuracies in the forecasts
are mainly due to background noise and the tidal component.

By contrast,
picks in quadrants III and IV (blue and black points)
correspond to linear combinations of two ${\bm g}_k$'s.
There is a model mismatch with the 1 and 4 protocols,
but not for the 2 and matched protocols
(note that the blue and black points in~(b) and~(d) are necessarily identical).
Interestingly,
when the mismatch is due to one less ${\bm g}_k$
in the 1 protocol,
there are highly structured deviations from the diagonal line.
On the other hand,
when the mismatch is due to two extraneous ${\bm g}_k$'s,
as happens with the 4 protocol,
the scatter about the diagonal line visually does not increase much
over that of the 2 protocol (no mismatch).
\begin{figure}
\hskip10pt
\includegraphics{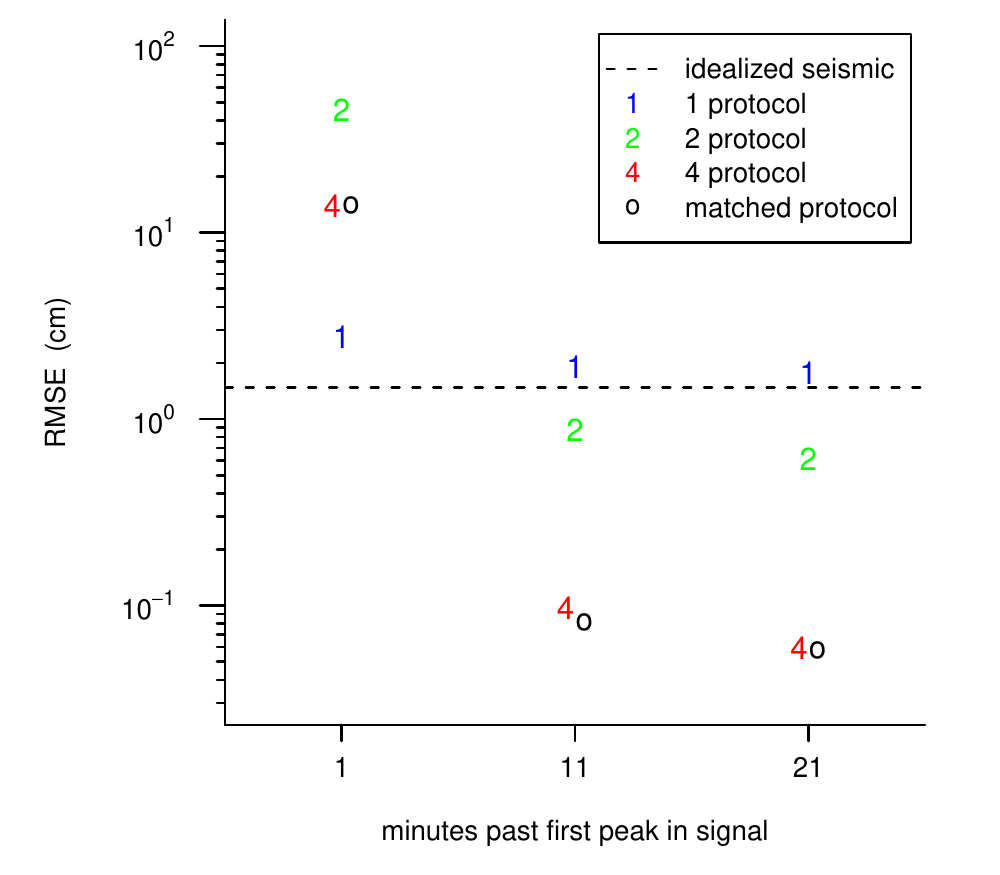}
\caption{
Root-mean-square errors (RMSEs) for forecasted versus presumed maximum
wave heights for 1000 realizations of representative triad
(central unit source {\tt ac026b}, buoy 46403 and impact site Hilo).
There are twelve RMSEs in all
(each computed as per Eq.~({\bl\ref{eq:RMSE}}))
corresponding
to selecting one of the four protocols
(the 1, 2, 4 and matched protocols)
in combination with three different amounts of data
(dictated by 1, 11 and 21 minutes past the first peak
in each constructed signal $\bm t$)
to be placed into $\bar {\bm y}$
in Eqs.~({\bl\ref{eq:BasicModel}}) and~({\bl\ref{eq:RegressionModel}}).
The RMSEs corresponding to the four scatterplots
of Fig.~{\bl\ref{fig:ScatterPlots}} are plotted
above the `21' label on the horizontal axis.
The horizontal dashed line indicates the RMSE
for forecasted heights based on idealized seismic information only
(and not any simulated data collected by buoy 46403)
}
\label{fig:RMSEsSingleBuoy}
\end{figure}

We can summarize how well forecasted maximum wave heights match
up with the presumed heights
by computing a root-mean-square error (RMSE) for each scatterplot.
Letting $\max\,\{\hat {\bm h}_l \}$ and $\max\,\{{\bm h}_l \}$
represent the forecasted and presumed heights for the $l$th realization,
this error is defined as
\begin{equation}\label{eq:RMSE}
\hbox{RMSE}
=
\left[\frac{1}{1000}\sum_{l=1}^{1000}\left(\max\,\{\hat {\bm h}_l \} - \max\,\{{\bm h}_l \}\right)^2\right]^{1/2}.
\end{equation}
Figure~{\bl\ref{fig:RMSEsSingleBuoy}} shows the RMSEs
for the four protocols in combination
with three settings $N$ for the amount of data used to estimate the source coefficients
(dictated by up to either 1, 11 or 21 minutes past the first peak in the signal $\bm {t}$).
The four RMSEs corresponding to the scatterplots in Fig.~{\bl\ref{fig:ScatterPlots}}
are shown above the `21' label on the horizontal axis.
In addition this figure has a dashed line showing the RMSE for a so-called seismic solution,
for which a forecast is generated using $4 {\bm g}_5$.
This procedure assumes idealized seismic information about the tsunami event,
namely, that
the generating earthquake occurs within the central unit source {\tt ac026b}
and that the magnitude of the earthquake suggests setting the source coefficient to 4,
which corresponds to our standard normalization $\sum_k \alpha_k = 4$
for the relocated unit source.
The seismic-based forecast is of interest
because it is arguably the best
that can achieved without any information supplied
by data from \DART\ buoy 46403.

For all four protocols for formulating the regression model,
the RMSEs decrease as more data are collected by the \DART\ buoy.
The decrease is minimal for the 1 protocol,
but substantial for the other three protocols
(in particular, there is a drop of more than two orders of magnitude
for the 4 and matched protocols).
The 1 protocol always involves a mismatch between the signal and its model.
For the other protocols, there is at least a 50\% chance of no mismatch.
The poor performance of the 1 protocol points out the need for an adequate regression model.
On the other hand, there is little difference in the RMSEs
for the 4 and matched protocols
even though the former involves mismatches in about 50\% of the realizations,
whereas the latter involves none.
In contrast to the 1 protocol,
the nature of the mismatch with the 4 protocol
is too many ${\bm g}_k$'s rather than too few.
This result suggests that
having extraneous ${\bm g}_k$'s in the model (overfitting)
does not significantly impact the forecasted wave heights.

The seismic-based forecast outperforms the buoy-based forecasts
either when there is an insufficient amount of data
or when the regression model is always inadequate
due to missing ${\bm g}_k$'s (the 1 protocol).
When there is too little data,
forecasts deteriorate considerably
as the complexity of the regression model increases
(i.e., more coefficients must be estimated).
For our representative triad,
there is some value to be gained by waiting
the extra 10 minutes from 11 minutes past the first peak
to 21 minutes (the RMSEs drop by 60\% to 70\%).
For other triads involving different central unit sources and different buoys,
the gain can be more substantial.

For the network assessment discussed in the next section,
we will concentrate on the 4 protocol
with $N$ selected as dictated by 21 minutes after the first peak in the signal.
This choice is close to the best RMSE for most -- but not all -- triads
and should reflect the ability of a particular buoy
in forecasting wave heights at an impact site
once enough data have been collected and once an adequate regression model
has been identified. 

\section{Network Assessment}\label{sec:network}

In the previous section,
we described an approach for assessing the effectiveness of a particular buoy
in forecasting wave heights in the open ocean outside of an impact site of interest
when an earthquake-generating tsunami arises
from within a particular central predesignated unit source.
The approach takes into account major factors
influencing the quality of the forecasts
(including tides, background noise, uncertainties in the signal model
and amount of available buoy data).
We concentrated on a representative triad
(unit source {\tt ac026b}, \DART\ buoy 46403 and impact site Hilo),
but considering a single triad is not enough to assess a network of \DART\ buoys.
In this section we turn our attention to network assessment.
To focus our discussion,
we consider how well the existing network of \DART\ buoys does
in responding to tsunami events originating
first from the Aleutian Islands (Sect.~{\bl\ref{subsec:AleutianIslands}})
and then from South America (Sect.~{\bl\ref{subsec:SouthAmerica}}).

\begin{figure}
\hskip-5pt
\includegraphics{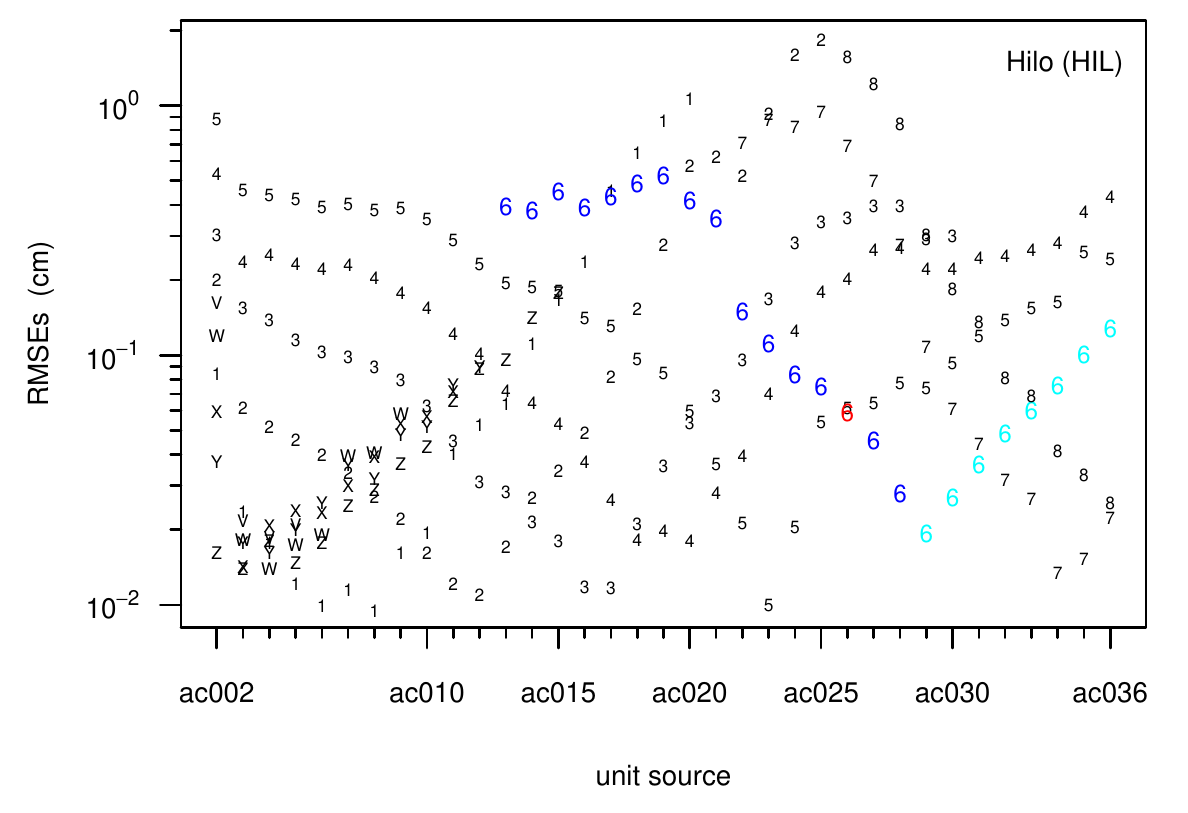}
\caption{
RMSEs for forecasted maximum wave heights in the open ocean outside of Hilo
based on 14 buoys (labeled as indicated in Fig.~{\bl\ref{fig:Aleutian}})
in combination with central unit sources {\tt ac002b} to {\tt ac036b}
from the Aleutian--Alaskan subduction zone.
The RMSEs for buoy 46403 (labeled as {\tt 6}) are colored,
with red indicating its pairing with {\tt ac026b}
}
\label{fig:summaryAlaska}
\end{figure}

\subsection{First Case Study: Aleutian Islands}\label{subsec:AleutianIslands}

Figure~{\bl\ref{fig:Aleutian}} shows the locations of 74 predesignated unit sources
tiling the Aleutian--Alaskan subduction zone,
the 14 \DART\ buoys closest to this zone
(8 along the Aleutian Islands (labeled {\tt 1} to {\tt 8}) and 6 to the west ({\tt U} to {\tt Z}))
and three impact sites (Hilo, Crescent City and Port San Luis --
these are labeled as HIL, CCY and PSL).
To evaluate this network of buoys,
we consider all possible triads involving one buoy,
one impact site and one central unit source selected from {\tt ac002b}, \dots, {\tt ac036b}
(there are thus $14\times3\times35 = 1470$ triads in all).
Figure~{\bl\ref{fig:summaryAlaska}} shows RMSEs
of forecasted maximum wave heights in the open ocean outside of Hilo.
These RMSEs are based on the 4 protocol
and use of 21 minutes of data after the first peak in the constructed tsunami signal.
The figure shows RMSEs only for those buoy/unit source combinations
such that, as called for by standard operating procedures,
a warning to Hilo could be issued at least three hours in advance of the arrival of the tsunami
(when this criterion is applied to all three impact sites,
the number of relevant triads shrinks from 1470 down to 835).
The RMSEs are plotted categorically versus unit source.
Those for \DART\ buoy 46403 (labeled as {\tt 6}) are colored,
with the RMSE for unit source {\tt ac026b} shown in red
(this RMSE is also displayed in Fig.~{\bl\ref{fig:RMSEsSingleBuoy}}
as the red {\tt 4} in the 21 minutes category).
The RMSEs for {\tt ac002b}, \dots, {\tt ac012b} are not shown
because a tsunami originating from one of these unit sources
would not arrive at 46403 soon enough to issue a timely warning to Hilo.
\begin{figure}
\hskip-15pt
\includegraphics{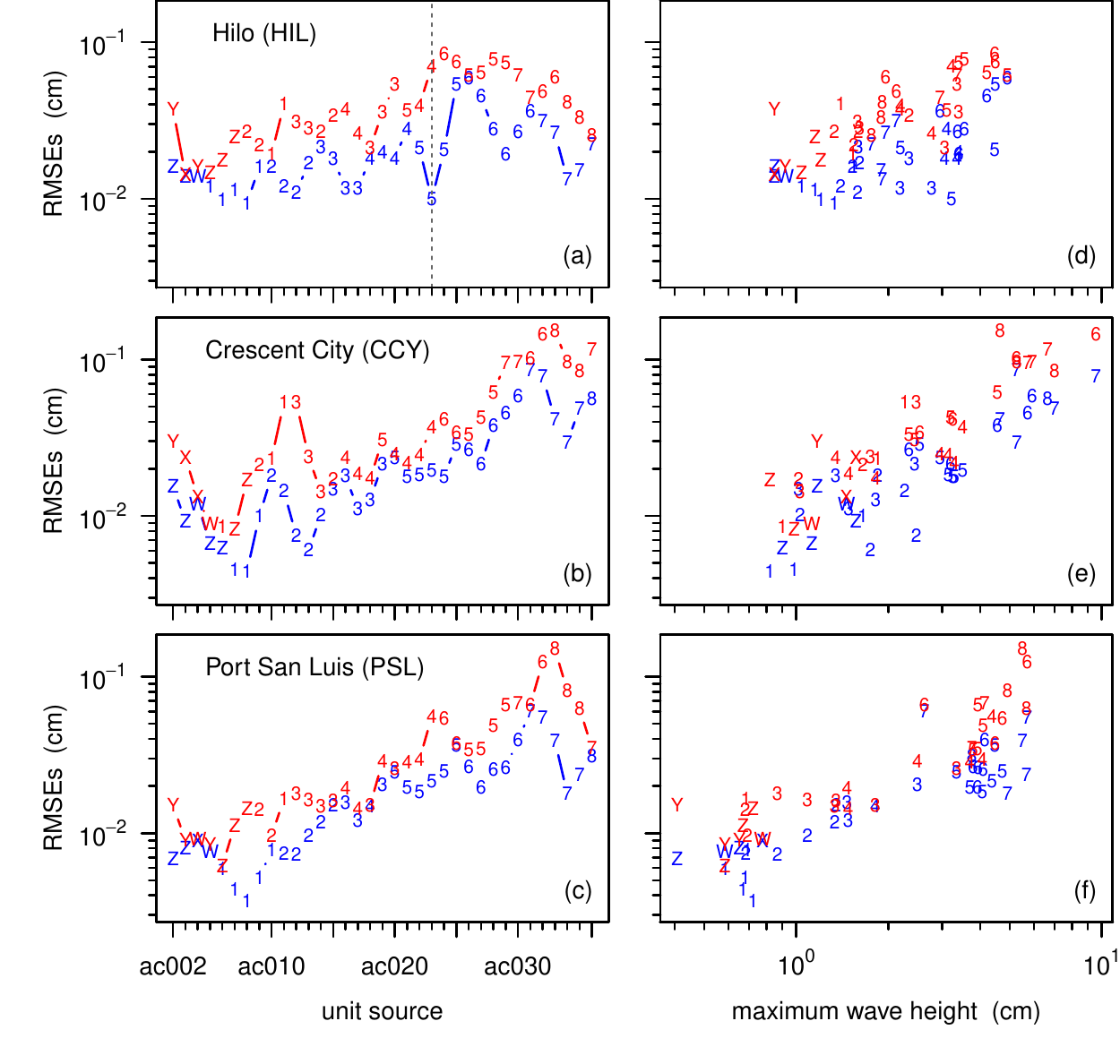}
\caption{
Best (blue) and second best (red) RMSEs for forecasted maximum wave heights in the open ocean
outside of, from top to bottom rows, Hilo, Crescent City and Port San Luis.
The RMSEs are plotted versus central unit sources {\tt ac002b}, \dots, {\tt ac036b} in the left-hand column and, in the right-hand, versus maximum wave heights.
Plot~(a) shows RMSEs extracted from Fig.~{\bl\ref{fig:summaryAlaska}},
and the vertical dashed line crosses the RMSEs for unit source {\tt ac023b}
in combination with buoys {\tt 5} and {\tt 4}
}
\label{fig:bottomLines3x2}
\end{figure}

For assessing the ability of a network of buoys to provide good forecasted maximum wave heights,
individual RMSEs are not as much interest as the best RMSE
that can be achieved across all buoys
for an event originating from within a particular unit source.
The second best RMSE is also of interest because
a comparison of this with the best RMSE
tells us how much degradation in forecasted height there would be
if the buoy with the best RMSE were to become inoperative.
Figure~{\bl\ref{fig:bottomLines3x2}(a)} simplifies Fig.~{\bl\ref{fig:summaryAlaska}}
by showing just the first (blue) and second best (red) RMSEs.
There is a prominent pinching pattern between the two sets of RMSEs.
The unit sources where the pinches occur are ones
for which the distances from the source to the two associated buoys are approximately equal.
For events originating from pinch locations,
if the buoy with the best RMSE were to drop out,
there would be little impact on the ability of the network
to forecast wave heights outside of Hilo
because of the existence of another buoy that performs almost as well.
Figures~{\bl\ref{fig:bottomLines3x2}(b)} and~(c)
are similar to~(a),
but now use Crescent City and Port San Luis as impact sites rather than Hilo.
Pinching patterns are in evidence again,
with the locations of the pinches occurring close to the same unit sources as for Hilo.
With some exceptions,
the first and second best RMSEs are also associated with the same buoys.
These results indicate that
there is some degree of commonality across impact sites
that are all distant from the unit sources.

In addition to pinching patterns,
there are particularly prominent upward trends in Fig.~{\bl\ref{fig:bottomLines3x2}(b)}
and~(c)
for Crescent City and Port San Luis
(the correlation of the displayed log RMSEs with unit source indices 2, \dots, 36 
is $0.78$ for Crescent City and $0.84$ for Port San Luis,
while Hilo has a smaller value of $0.52$).
A glance at Fig.~{\bl\ref{fig:Aleutian}} shows that
Hilo is more centrally located 
with respect to the Aleutian--Alaskan unit sources
than the California sites, both of which are off to one side.
The travel times from the 35 central unit sources to the California sites
both have a nearly monotonic decay proceeding from {\tt ac002b} to {\tt ac036b},
whereas those for Hilo decrease from {\tt ac002b} to {\tt ac021b}
and then increase along the remaining 15 unit sources.
As travel time increases, the maximum height of a tsunami signal should tend to decrease.
Presuming an inverse relationship between travel time and signal height,
the prominent upward trends
at the California sites
suggest that RMSE increases as signal height increases.
To explore this suggestion,
the right-hand column of Fig.~{\bl\ref{fig:bottomLines3x2}} shows
the RMSEs of the forecasted maximum wave heights for each of the impact sites
versus the maximum wave heights outside of the three impact sites.
The maximum wave heights were extracted from a propagation data base of predictions
linking the center of each central unit source 
with the open ocean location outside of the impact site.
Crescent City and Port San Luis have prominent upward trends, but Hilo, less so. 
The correlation of log RMSEs with log maximum wave heights
is $0.58$ for Hilo, $0.83$ for Crescent City and $0.84$ for Port San Luis.
These correlations agree well with the corresponding ones from
the left-hand column of Fig.~{\bl\ref{fig:bottomLines3x2}},
thus offering support for the hypothesis
that the upward trends in that column 
are due at least in part to RMSE and signal height being positively related.
\begin{figure}
\hskip-10pt
\includegraphics{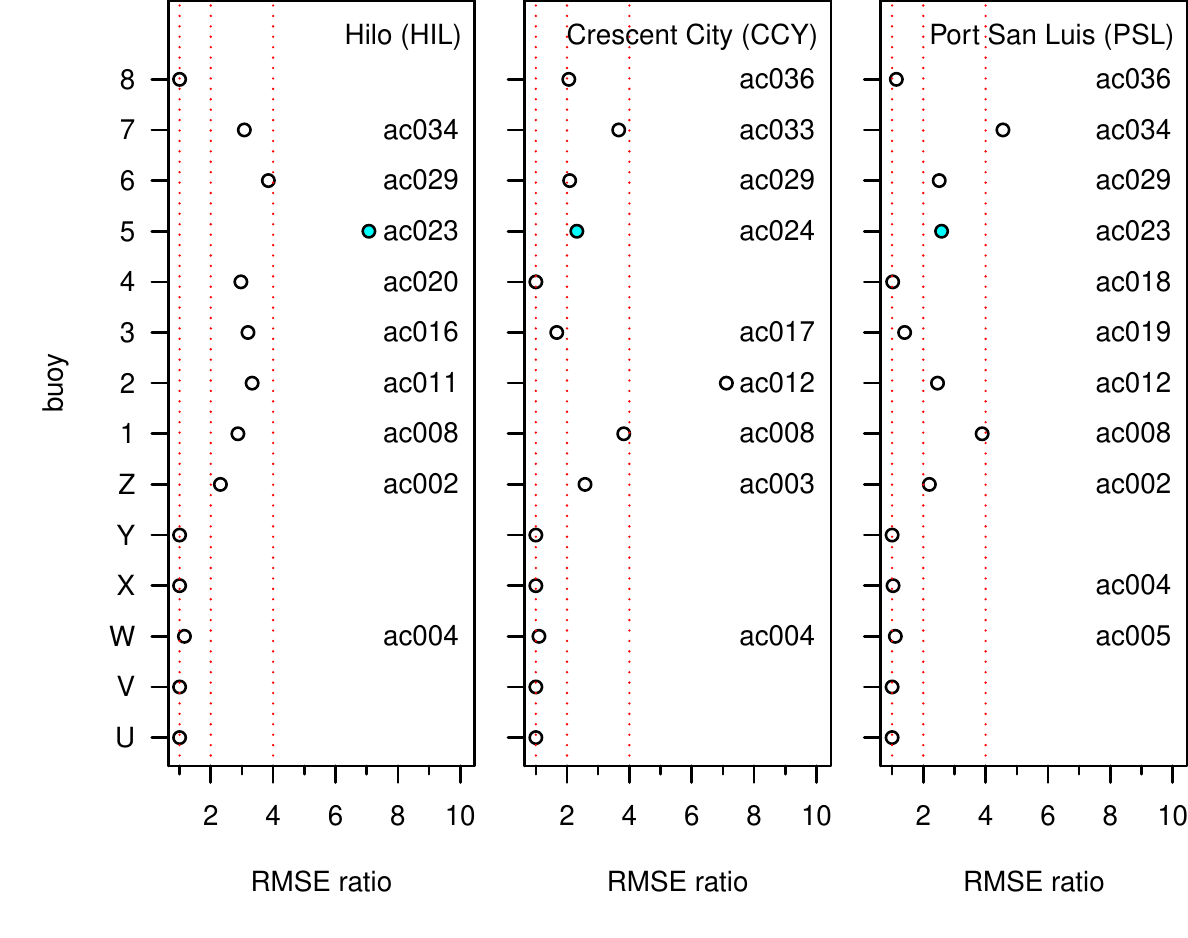}
\caption{
Effect on performance of \DART\ buoy network
across 35 unit sources and at three impact sites
due to one-by-one removal of the 14 buoys
depicted in Fig.~{\bl\ref{fig:Aleutian}}.
For each buoy, the largest increase in RMSE over the 35 unit sources is shown 
relative to the best available RMSE when that buoy is dropped as compared to
when the full network of buoys is available.
The case where buoy {\tt 5} is dropped is highlighted with colored circles.
The one corresponding to Hilo is the change in RMSE
indicated by the vertical dashed line in Fig.~{\bl\ref{fig:bottomLines3x2}(a)}.
The three red dashed lines mark RMSE ratios of 1, 2 and 4.
These ratios are used to define a green, yellow and red network status --
see Fig.~{\bl\ref{fig:GreenYellowRed}} and the discussion pertaining to it
}
\label{fig:leaveOneOut}
\end{figure}

To quantify how well the existing network of \DART\ buoys
can respond to events emanating from the Aleutian--Alaskan subduction zone,
let us focus on forecasting maximum wave heights based on single buoys
and consider what happens if one buoy were to drop out.
As a measure of both the contribution of individual buoys to the network
and the overall strength of the network,
consider the maximum relative increase in RMSE
across all 35 central unit sources due to a particular buoy becoming inoperative.
Suppose, for example, that buoy {\tt 5} becomes inoperative.
Focusing on Hilo,
Fig.~{\bl\ref{fig:bottomLines3x2}(a)} indicates
that, if a tsunami were to emanate from unit source {\tt ac023b}
(corresponding to the dashed vertical line),
loss of {\tt 5} would result in a 7 fold increase in the best available RMSE
because we would then have to rely on buoy {\tt 4}
(the RMSEs would also increase at the California sites, but less so). 
For Hilo, buoy {\tt 5} also has the best RMSE for three additional unit sources
({\tt ac022b}, {\tt ac024b} and {\tt ac025b}),
but the relative increases in RMSE due to {\tt 5} being gone
are all smaller than that for {\tt ac023b}.
For events emanating from any of the remaining 31 unit sources, 
the loss of {\tt 5} would not cause a deterioration in the ability of the network
to forecast Hilo's wave heights.

Figure~{\bl\ref{fig:leaveOneOut}} quantifies
how the network is affected when a single buoy becomes inoperative.
For every buoy that can be used to deliver a warning at least three hours in advance
for a particular impact site due to an event arising from at least one of the 35 central unit sources,
we examine the ratio of two particular RMSEs for each unit source.
The first RMSE is the best available RMSE after elimination of the buoy in question,
and the second is the best RMSE when all \DART\ buoys are operational.
We then concentrate on just the largest such ratio across all unit sources.
The left panel of Fig.~{\bl\ref{fig:leaveOneOut}} shows these RMSE ratios for 14 buoys and at Hilo.
The smallest possible ratio is unity, which is indicated by the left-most red dashed line.
A ratio of unity occurs when a buoy's RMSEs are never the best at any of the 35 unit sources.
Such a buoy does not participate in the blue portion of Fig.~{\bl\ref{fig:bottomLines3x2}(a)}
and typically is not in close proximity to at least one unit source
(see Fig.~{\bl\ref{fig:Aleutian}}).
Under a worst case scenario,
loss of these buoys would not degrade any forecasts.
By this measure,
buoys with an RMSE ratio of unity are of secondary importance to Hilo
for responding to events arising from the Aleutian Islands
(they are primarily intended to handle events arising elsewhere).
When the ratio is greater than unity,
the network deteriorates at least to some degree 
when the buoy in question is out of commission.
For these buoys, Fig.~{\bl\ref{fig:leaveOneOut}} indicates
the unit source linked to the largest proportional increase in RMSE.
The largest RMSE ratio (indicated by a colored solid circle) occurs
when buoy {\tt 5} is not available to handle an event arising from unit source {\tt ac023b}.
Thus, assuming a worst case scenario,
loss of buoy {\tt 5} would cause the largest degradation in the network's ability
to forecast maximum wave heights at Hilo.
The middle and right panels of Fig.~{\bl\ref{fig:leaveOneOut}} show
corresponding results for Crescent City and Port San Luis.
For a buoy whose ratios are greater than unity for multiple impact sites,
the associated unit sources are either the same or quite close to one another.
With the exceptions of buoy {\tt 2},
the RMSE ratios for the two California sites are remarkably similar.

\begin{figure}
\hskip-10pt
\includegraphics{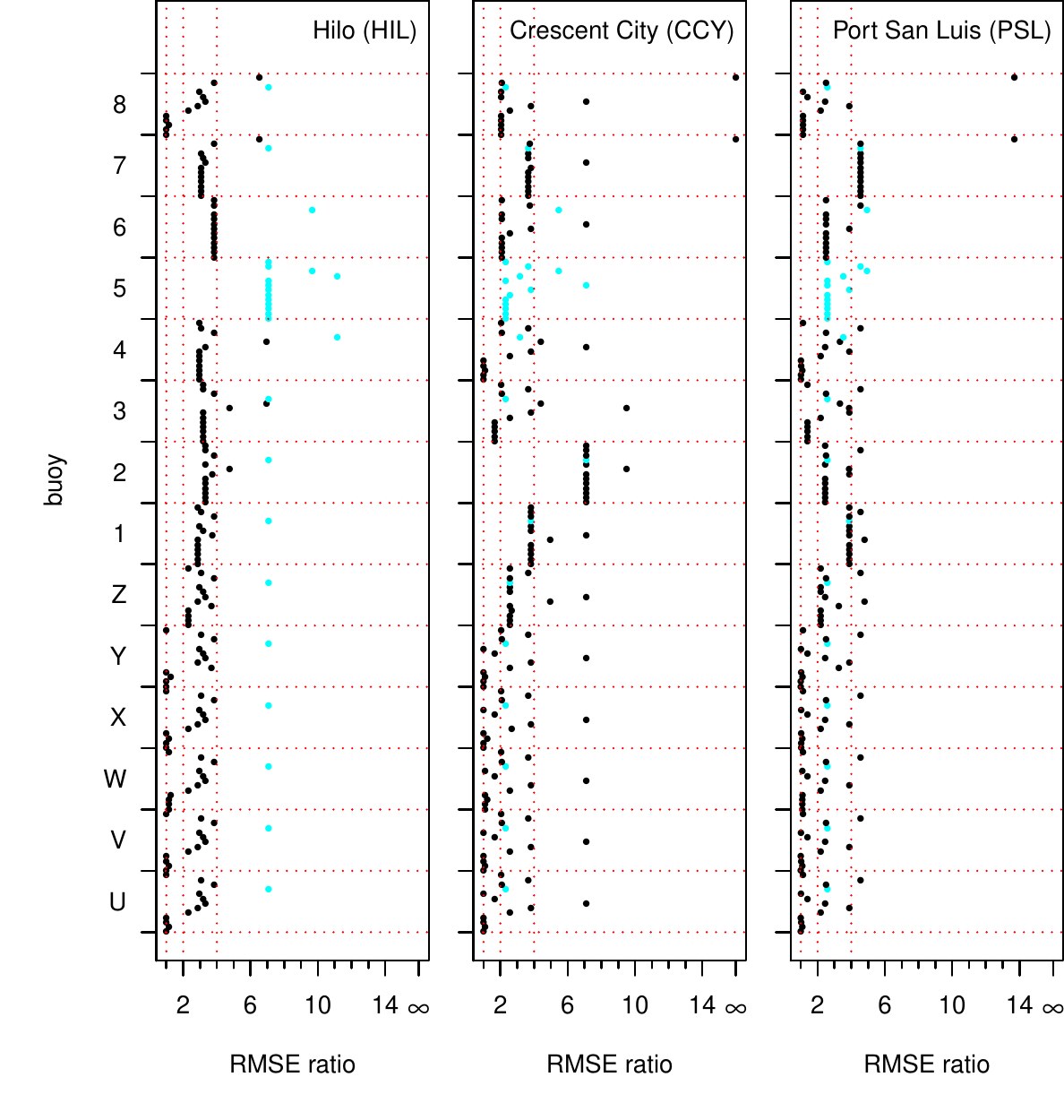}
\caption{
Effect on performance of \DART\ buoy network
across 35 unit sources and at three impact sites
due to concurrent dropout of two of 14 buoys depicted in Fig.~{\bl\ref{fig:Aleutian}}.
For each pair of buoys, the largest increase in RMSE over the 35 unit sources is shown 
relative to the best available RMSE when both buoys are dropped
as compared to when the full network of buoys is available.
Any cases where buoy {\tt 5} is one of the dropped buoys
are highlighted with colored points.
The three red dashed lines mark RMSE ratios of 1, 2 and 4.
These ratios are used to define a green, yellow and red network status --
see Fig.~{\bl\ref{fig:GreenYellowRed}} and the discussion pertaining to it
}
\label{fig:leaveTwoOut}
\end{figure}

Figure~{\bl\ref{fig:leaveTwoOut}} explores the effect on network performance
when two buoys drop out.
As in Fig.~{\bl\ref{fig:leaveOneOut}},
there are three panels, one for each impact site.
Each panel has 15 red dotted horizontal lines,
which breaks the panel up into 14 subplots,
one for each of the relevant buoys.
Consider the bottommost subplot in the left-hand panel,
which is for buoy {\tt U} in combination with Hilo.
There are 13 dots in the subplot,
each indicating the maximum ratio of two particular RMSEs across the 35 unit sources.
The first RMSE is the best remaining after {\tt U} drops out along with one of the other 13 buoys,
and the second is the best RMSE when all 14 buoys are present
(going from top to bottom in the subplot, the dots are associated with buoys {\tt 8} to {\tt V}).
The colored dot is the RMSE ratio when both {\tt U} and {\tt 5} are inoperative
(the colored dots in the other subplots also indicate ratios in which {\tt 5} is involved).
Because buoy {\tt U} never has the best RMSE for any of the 35 unit sources,
the ratios merely reflect what happens when the second buoy drops out.
As a result, the dots in the subplot have same pattern as the topmost 13 circles
in the left-hand panel of Fig.~{\bl\ref{fig:leaveOneOut}}.
Thus, when examining any of the other 13 subplots,
we can compare it to the bottommost subplot
to ascertain of how much the leave-out-two pattern deviates from the leave-one-out pattern.
The same holds for the middle and right-hand panels for Crescent City and Port San Luis.

The subplots for buoys {\tt V} to {\tt 8} in Fig.~{\bl\ref{fig:leaveTwoOut}}
are organized in a manner similar to that for {\tt U}.
Consider the subplot for buoy {\tt 5} in combination with Hilo.
Because any ratio involving {\tt 5} is represented by a colored dot,
all of the dots in this subplot are colored
(in each of the other 13 subplots for Hilo, there is exactly one colored dot).
With two exceptions, the ratios are all identical to 7.1.
The identical ratios come about
because dropping {\tt 5} leads to the largest proportional increase in RMSE
across all 35 unit sources
(see Fig.~{\bl\ref{fig:bottomLines3x2}(a)}). 
The two exceptional ratios are 9.7 and 11.2.
A dot in a given subplot representing the RMSE ratio for a particular pair of buoys 
must be replicated in one other subplot.
The dot for 9.7 is replicated in the subplot for {\tt 6},
and the one for 11.2 has a duplicate in the subplot for {\tt 4}.
Hence, under the worse case scenario,
the concurrent loss of {\tt 5} and {\tt 4}
would lead to an order of magnitude increase in RMSE.
This scenario involves an tsunami-generating earthquake occurring in unit source {\tt ac023b},
for which the third best RMSE is for buoy {\tt 6}.
The second worst case scenario would be to lose {\tt 5} and {\tt 6}
for an event in {\tt ac028b},
for which the third best RMSE is for buoy {\tt 4}.

The subplots for eight buoys ({\tt Z} to {\tt 7})
in the left-hand panel of Fig.~{\bl\ref{fig:leaveTwoOut}}
all have ratios greater than unity.
These same buoys also have ratios greater than unity
under the leave-one-out scenario
(left-hand panel of Fig.~{\bl\ref{fig:leaveOneOut}}).
If one of these buoys were to be inoperative,
loss of a second buoy from this set of eight
could further compromise the network's ability to forecast maximum wave heights at Hilo.
The maximum ratios in all eight subpanels are associated with blue dots,
which means,
once any buoy amongst these eight other than {\tt 5} becomes inoperative,
the worst case scenario is to lose {\tt 5} also
(as noted previously, the worse case overall is the loss of {\tt 5} and {\tt 4} together).
If any of the 14 buoys were to becomes inoperative,
a study of the corresponding subplot would tell us
how much the ability of the network to forecast Hilo wave heights would deteriorate
if another buoy were to become inoperative before the first one is returned
to operational status.

The middle and right-hand panels of Fig.~{\bl\ref{fig:leaveTwoOut}}
show leave-two-out RMSE ratios for the two California impact sites.
For Crescent City (middle panel), the worst case scenario occurs
if buoys {\tt 7} and {\tt 8} were inoperative
during an event originating in either unit source
{\tt ac034b}, {\tt ac035b} or {\tt ac036b}
(see Fig.~{\bl\ref{fig:bottomLines3x2}(b)}).
These two buoys are the only ones
that can provide a warning three hours in advance to Crescent City.
Because there is no other buoy to rely on,
we can consider the RMSE ratio to be infinite.
The second worst case scenario occurs when buoys {\tt 2} and {\tt 3} drop out
for an event from {\tt ac012b} leading to the RMSE ratio of 9.5,
with {\tt 1} providing backup.
After these worse cases,
there are a number of RMSE ratios equal to 7.1,
all of which involve buoy {\tt 8} and unit source {\tt ac012b}.

Turning now to Port San Luis (right-hand panel of Fig.~{\bl\ref{fig:leaveTwoOut}}),
the RMSE for the worst case scenario is 13.7,
which occurs when {\tt 7} and {\tt 8} were inoperative
and is associated with {\tt ac034b}
(see Fig.~{\bl\ref{fig:bottomLines3x2}(c)}).
With these two buoys gone,
buoy {\tt 6} has the best RMSE amongst the remaining buoys
satisfying the three hour warning time constraint.
The second and third largest RMSE ratios are 4.9 and 4.8.
The latter is of interest because
the two buoys involved are {\tt 1} and {\tt Z},
and the unit source is {\tt ac008b}
(this is in agreement with Fig.~{\bl\ref{fig:bottomLines3x2}(c)}).
With both of these buoys gone,
buoy {\tt W} has the best remaining RMSE.
Note that, in this case, 
the RMSE-based metric for network evaluation differs from one based solely on travel time
since the travel times between unit source {\tt ac008b} and
buoys {\tt X} and {\tt Y}
are both shorter than that for {\tt W}
(see Fig.~{\bl\ref{fig:Aleutian}}).

We can also explore the effect on network performance when three buoys drop out.
For Hilo, the worse case scenario is for buoys {\tt 4}, {\tt 3} and {\tt 5}
to be unavailable when an event arises from unit source {\tt ac020b}.
Buoy {\tt 6} then has the best RMSE amongst the remaining buoys
satisfying the three hour warning time constraint,
and the associated RMSE ratio raises to 23.0
(as compared to 11.2 when two buoys -- {\tt 5} and {\tt 4} -- drop out
during an event arising from unit source {\tt ac023b}). 
For Crescent City,
we have already noted that loss of {\tt 7} and {\tt 8}
means that none of the remaining buoys can provide a three hour warning
for events from either {\tt ac034b}, {\tt ac035b} or {\tt ac036b}.
With the additional loss of {\tt 6},
events from {\tt ac031b}, {\tt ac032b} or {\tt ac033b}
are also problematic because of lack of a fourth buoy
that can provide a three hour warning.
There are similar concerns for {\tt ac029b} and {\tt ac030b}
when {\tt 5}, {\tt 6} and {\tt 7} are jointly inoperative.
Turning to Port San Luis,
the worst cases scenario involves buoys {\tt 7}, {\tt 8} and {\tt 6};
the associated unit source is {\tt ac034b};
and backup comes from buoy {\tt 5} with an RMSE ratio of 42.2.
The buoys and unit sources match up well with the leave-two-out case.

\begin{figure}
\hskip10pt
\includegraphics{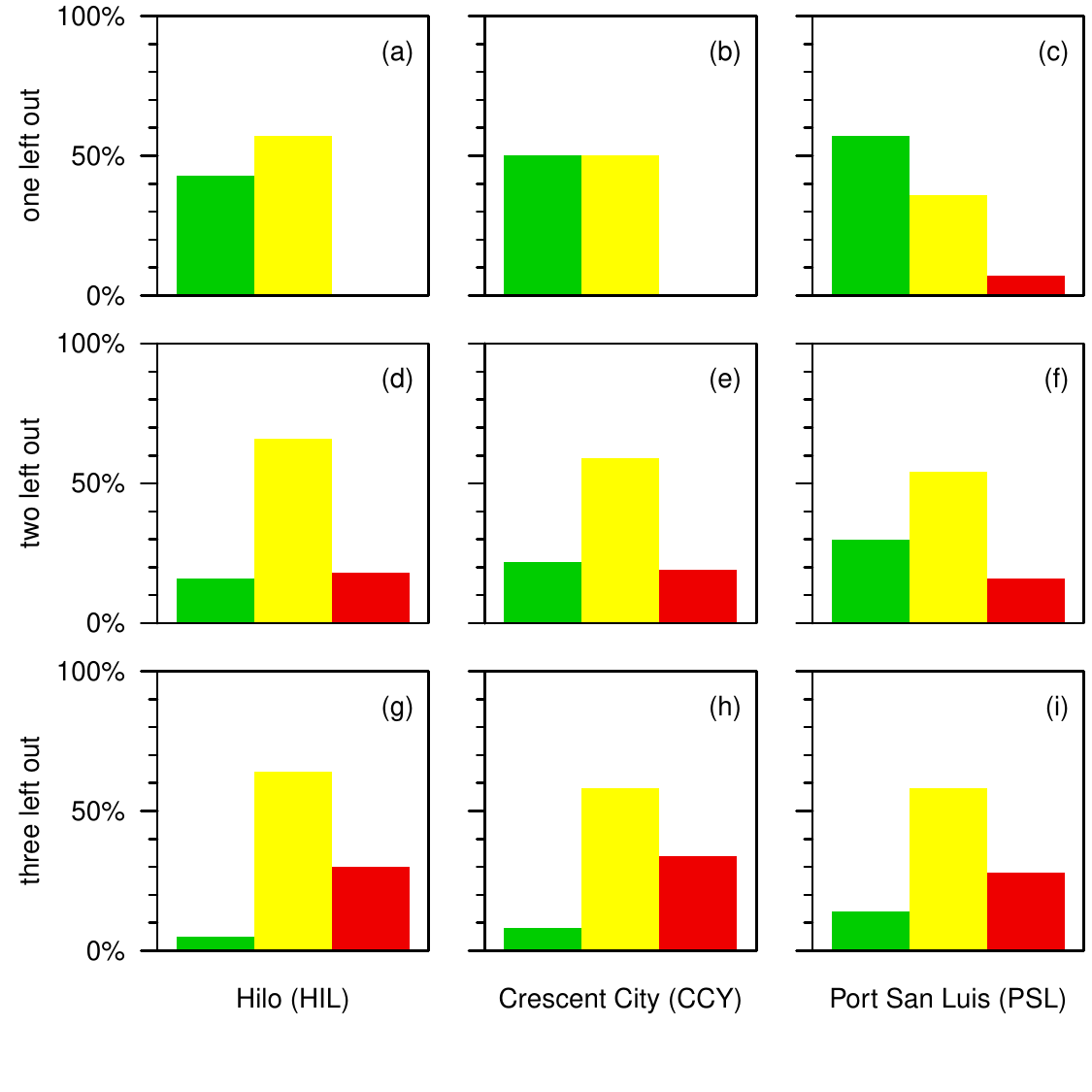}
\caption{
Percentages of cases that would result in a green, yellow or red network status
for Hilo, Crescent City and Port San Luis (left to right columns)
due to 1, 2 or 3 of 14 \DART\ buoys becoming inoperative (top to bottom rows, respectively)
}
\label{fig:GreenYellowRed}
\end{figure}

To help assess the effect of one, two or three buoys becoming inoperative,
we introduce a simple indicator of the current status of the network
reflecting its ability to forecast maximum wave heights at a given impact site.
The indicator has three levels:
\begin{enumerate}
\item if the RMSE ratio is less than 2,
we award the network a `green' status (the network is healthy); 
\item a ratio between 2 and 4 indicates a `yellow' status (the network has deteriorated somewhat); and 
\item a ratio greater than an 4 is a `red' status (there is serious deterioration) 
\end{enumerate}
(the values 2 and 4 are placeholders for our demonstration,
but are arbitrary and subject to change when used in practice).
In Fig.~{\bl\ref{fig:leaveOneOut}},
there are vertical dotted lines at 2 and 4 in each of the plots --
these delineate the boundaries between the three levels.
Considering the plot for Hilo,
we see that the network would still have a green status
even if any one of 6 buoys ({\tt U} to {\tt Y} or {\tt 8}) out of a total of 14 (i.e., 43\%)
were to drop out by itself;
on other other hand, losing any one of {\tt Z} or {\tt 1} to {\tt 7} would result
in a yellow status (57\%).
Figure~{\bl\ref{fig:GreenYellowRed}}(a)
depicts these percentages,
and plots~(b) and~(c) show corresponding results for the two California sites.
The middle and bottom rows of Fig.~{\bl\ref{fig:GreenYellowRed}}
shows the percentages when two or three buoys drop out together.
Not surprisingly,
as more buoys drop out,
the percentage of cases with red status increases
for a given impact site.
Note that three buoys can drop out
but there can still be a green rating
at all three impact sites for events arising from Aleutian Islands
(but presumably not elsewhere).
On the other hand,
dropout of the single buoy {\tt 7} can raise a red alarm for Port San Luis.
Once a particular buoy has dropped out, we can assess the effect of loss of one of the other buoys,
thus giving an indication of the urgency of returning the buoy to normal operation.
This exercise would address the question of how close the array is to falling into a red status.

As a concluding example,
the National Data Buoy Center ({\tt ndbc.noaa.gov}) reported
that four of the 14 buoys we have been considering
were inoperative at the time of writing (early June 2016),
namely, {\tt U}, {\tt V}, {\tt W} and {\tt 2}.
Loss of {\tt 2} alone is enough to raise a yellow alarm for Hilo and Port San Luis
and a red alarm for Crescent City (see Fig.~{\bl\ref{fig:leaveOneOut}}).
Additional loss of {\tt U}, {\tt V} and {\tt W} does not increase the alarm level
for Hilo or Port San Luis.
With these four buoys gone,
the alarm status for Hilo would switch from yellow to red
if either buoy {\tt 3} or {\tt 5} were to become inoperative,
whereas, for Port San Luis, only the additional loss of buoy {\tt 7}
would increase the alarm level. 

\subsection{Second Case Study: South America}\label{subsec:SouthAmerica}

In the preceding case study
we concentrated on evaluating an existing network of \DART\ buoys.
Here we illustrate another use for our proposed methodology,
namely, to assess the impact of extending an existing network
by adding a new buoy.

\begin{figure}
\centering
\includegraphics[width=4.5in,angle=0]{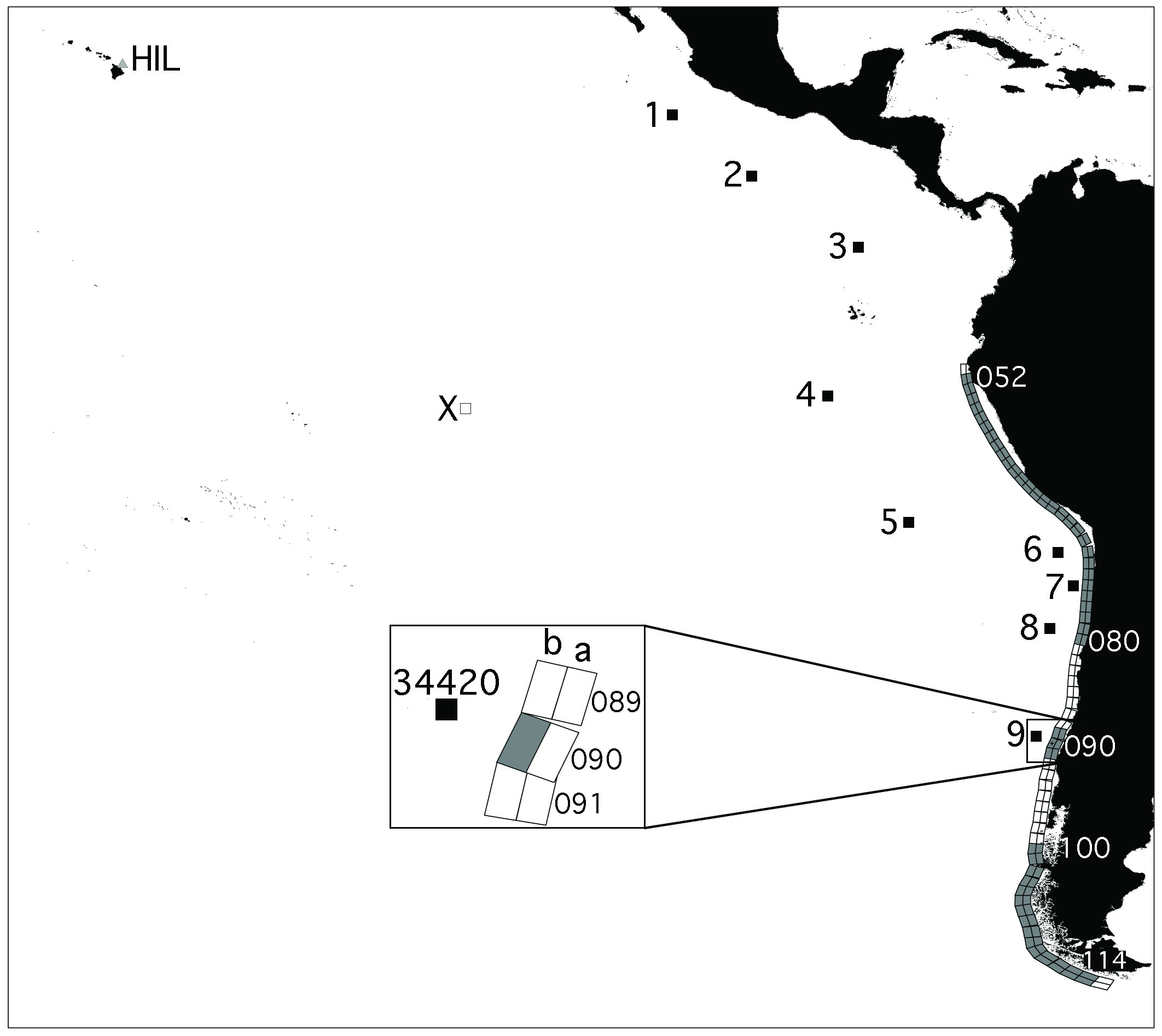}
\caption{
Locations of 130 unit sources along the West Coast of South America,
ten \DART\ buoys (squares labeled by {\tt 1} to {\tt 9} and {\tt X})
and the Hilo impact site (HIL).
The unit sources are arranged in two rows denoted by {\tt a} (eastern) and {\tt b} (western)
and in 65 columns denoted by, from top to bottom, {\tt 051} to {\tt 115}.
Going clockwise from the top,
the World Meteorological Organization designations for the 10 buoys are
({\tt 1})~43412,
({\tt 2})~43413,
({\tt 3})~32411,
({\tt 4})~32413,
({\tt 5})~32412,
({\tt 6})~32401,
({\tt 7})~32403,
({\tt 8})~32402,
({\tt 9})~34420 and
({\tt X})~51406.
The expanded view shows central unit source {\tt cs090b} and the five unit sources
surrounding it, along with the nearby buoy 34420
}
\label{fig:SouthAmerica}
\end{figure}

Figure~{\bl\ref{fig:SouthAmerica}} is a map of the West Coast of South America
showing the locations of 130 predesignated unit sources,
the Hilo impact site (HIL), nine operational \DART\ buoys (labeled {\tt 1} to {\tt 9})
and one decommissioned buoy ({\tt X}).
These buoys are mainly intended to provide forecasts for Southern Hemisphere impact sites.
Merely to exercise our proposed methodology,
we presume buoys {\tt 1} to {\tt 8} to be a preexisting network
and consider the benefit to Hilo of adding buoy {\tt 9}
(we do not report on results for Crescent City and Port San Luis
because they are quite similar to those for Hilo).
In addition we explore a hypothetical question:
what would have been the effect on Hilo
if, rather than augmenting the network by adding buoy {\tt 9} (a coastal buoy),
buoy {\tt X} (an open ocean buoy) had been reactivated instead?

\begin{figure}
\centering
\includegraphics[width=4.5in]{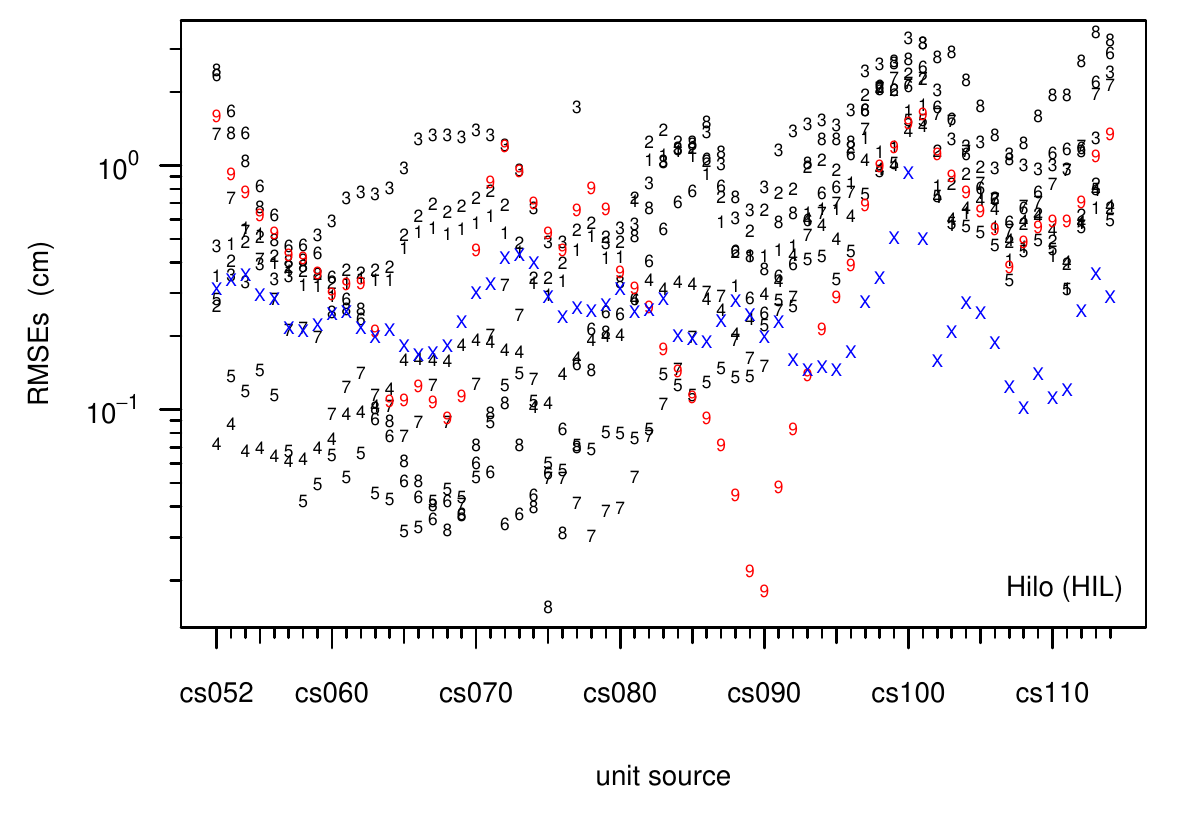}
\caption{
RMSEs for forecasted maximum wave heights in the open ocean outside of Hilo
based on ten buoys (labeled as indicated in Fig.~{\bl\ref{fig:SouthAmerica}})
in combination with central unit sources {\tt cs052b} to {\tt cs114b}
along the West Coast of South America
}
\label{fig:HiloSA}
\end{figure}

Figure~{\bl\ref{fig:HiloSA}} shows RMSEs of forecasted maximum wave heights
at Hilo for tsunami events arising from South America. 
In the same manner as Fig.~{\bl\ref{fig:summaryAlaska}}
for the Aleutian--Alaskan subduction zone,
these RMSEs are based on the 4 protocol and 21 minutes of data
after the first peak in the constructed tsunami signal.
The requirement that a warning to Hilo must be issued at least three hours in advance
of the arrival of the tsunami
is satisfied for all buoy/unit source combinations
with two exceptions:  buoy {\tt 1} with {\tt cs109b},
and the same buoy with {\tt cs114b}.
The RMSEs for buoys {\tt 9} and {\tt X} are colored, respectively, red and blue. 

For the network consisting of buoys {\tt 1} to {\tt 8}
(black symbols in Fig.~{\bl\ref{fig:HiloSA}}),
there is an increase in RMSE of approximately an order of magnitude
going from the northern-most central unit sources ({\tt cs052b} to {\tt cs080b})
to the southern-most ({\tt cs100b} to {\tt cs114b}).
This pattern is in contrast to Fig.~{\bl\ref{fig:summaryAlaska}}
for the Aleutian--Alaskan subduction zone,
where the RMSEs are not strikingly different across all unit sources.
Augmenting the network of eight buoys with coastal buoy {\tt 9} 
leads to a decrease in best RMSE by a factor to two or more
for unit sources {\tt cs087b} to {\tt cs094b}
(the unit source closest to {\tt 9} is {\tt cs090b}
-- see Fig.~{\bl\ref{fig:SouthAmerica}});
however, although there is some improvement in best RMSE
from unit source {\tt cs085b} up to {\tt cs097b},
there is none elsewhere.
The augmented network thus does not improve forecasts of wave heights at Hilo
for events arising from the southern-most unit sources.

Suppose now that we augment the network {\tt 1} to {\tt 8}
by adding the open ocean buoy {\tt X} rather than buoy {\tt 9}.
We would now have a factor of two or more decrease in best RMSE
for unit sources {\tt cs093b} to {\tt cs114b}
({\tt cs100b} is an exception -- there is still a noticeable decrease,
but less than a factor of two; in addition,
there are small improvements in best RMSEs for unit sources {\tt cs091b} and {\tt cs092b}).
Thus, in contrast to augmenting the network {\tt 1} to {\tt 8} with buoy {\tt 9},
augmenting with {\tt X} improves the response of the network in handling events events
from the southern-most unit sources.
The {\tt X} augmentation improves the best RMSE at 24 unit sources,
while the {\tt 9} augmentation, at 12 unit sources;
however, the drastic improvement offered by the {\tt 9} augmentation
at unit sources close to {\tt cs090} is not matched anywhere by the {\tt X} augmentation.

In terms of improving forecasts for Hilo,
augmenting the network {\tt 1} to {\tt 8} by reactivating the open ocean buoy {\tt X}
is arguably preferable to going with coastal buoy {\tt 9};
however, there are many other impact site of importance,
each of which would need to be evaluated in a similar manner,
and the individual evaluations would need to be combined
to come up with an objective evaluation of which buoy is the preferable augmentation
(how best to combine the evaluations is a subject for future research).
Our intent is not to take issue with the current placement of South American buoys,
but rather to demonstrate how our methodology can be used to evaluate
variations of an existing network.

\newpage 
\section{Discussion}\label{sec:discussion}
 
Four aspects of our method for evaluating buoy networks merit additional discussion.
First, as we have mentioned in passing,  
our approach is based on certain simplifications,
including (but not limited to)
\begin{enumerate}
\item tsunami-generating events that come from within a single unit source and
\item use of data from a single \DART\ buoy to forecast wave heights at an impact site
even though tsunami events are routinely observed at multiple buoys
(i.e., we do not combine together data
from several buoys to generate forecasts as is done during actual tsunami events).
\end{enumerate}
With regard to 1,
we are in effect assuming that highly localized tsunami events are useful
for assessing the effectiveness of a network of buoys.
With regard to 2,
we are assuming that forecasts based on each buoy separately can assess
the effectiveness of the overall network and that
using two or more buoys simultaneously won't lead to a significantly different evaluation.
While use of events originating from regions spanning more than a single unit source
and use of data from multiple buoys for generating forecasts
are more typical operationally,
going beyond simplifying assumptions 1 and 2 introduces
levels of complexity that would complicate the simulation procedures
needed for the network evaluation.
In addition, assumption 2 provides a way of evaluating the contribution
to the network of each buoy individually, which is more difficult to do
when buoys are used jointly for creating forecasts.

\begin{figure}
\hskip-15pt
\includegraphics{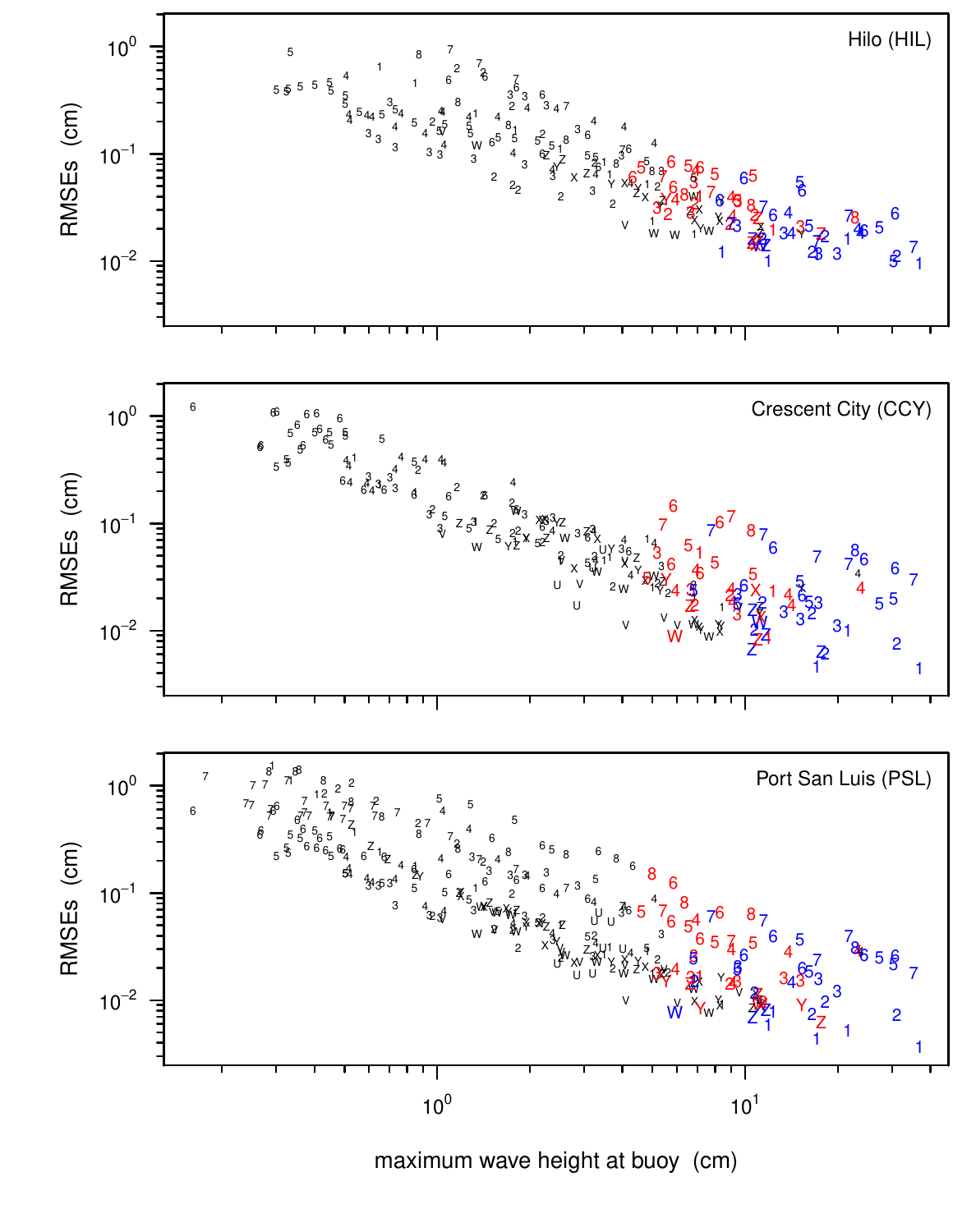}
\caption{
As in the right-hand column of Fig.~{\bl\ref{fig:bottomLines3x2}},
but now with the RMSEs of the maximum wave heights forecasted
at the open ocean outside of the impact sites
plotted versus maximum wave heights at the \DART\ buoys
rather than plotted versus maximum wave heights in the open ocean outside
of the impact sites.
While only the best RMSEs (colored by blue symbols) and second best (red)
are shown in Fig.~{\bl\ref{fig:bottomLines3x2}},
both  these and all remaining RMSEs (black) are shown here}
\label{fig:summaryAlaska35usWhBestStacked}
\end{figure}

Second,
our performance measure is predicated on tsunami events
coming from a specific subduction zone (e.g., Aleutian--Alaskan)
and heading toward a specific impact site (e.g., Hilo).
We have not addressed the question of how to take
the measures for different subduction zones and different impact sites 
and combine them together.
For example, combining measures for multiple subduction zones and Hilo
would lead to an evaluation of how well Hilo is serviced by the global buoy network.
On the other hand, combining measures for the Aleutian--Alaskan subduction zone
and multiple impact sites
would tell us how well the buoy network offers protection globally
for events arising from that particular subduction zone.
Combined measures such as these are certainly of interest,
but how best to do so is outside the scope of this article.

Third,
we have concentrated on impact sites that are distant from
the unit sources generating the tsunami events.
For these sites insisting on a 3 hr warning time is appropriate.
For sites that are close to a particular unit source
(of which there are many along the West Coast of South America),
a more stringent warning time (e.g., 30 min) must unfortunately be entertained.
A mixture of impact sites -- some distant and some close to a given unit source --
adds a degree of complexity (beyond the scope of this article) 
for assessing how well a tsunami event arising from such a source
is handled by a buoy network.

Fourth,
our work points to an interesting potential practical method
for quantifying uncertainties in forecasts.
Figure~{\bl\ref{fig:summaryAlaska35usWhBestStacked}} shows RMSEs at the three impact sites
versus maximum wave height at the buoys.
For a given buoy,
Fig.~{\bl\ref{fig:summaryAlaska35usWhBestStacked}} shows
that the lowest RMSEs are associated with the highest maximum wave heights. 
That is, buoys that are positioned well to sample the strongest portion of the main tsunami beam
yield the best forecasts in terms of RMSEs.
The relationship between the maximum wave heights at the buoys and the RMSEs
is roughly linear on a log/log scale, with a fairly strong sample correlation of $-0.88$.
This suggests that the RMSE in forecasts at impact sites
can be usefully assessed from the maximum wave height at the buoys.
This opens up the possibility of assigning a quality measure to forecasts
at impact sites based on a simple summary of the data collected at the buoy.   

\section{Summary and Conclusions}\label{sec:conclusions}

We have developed a performance measure that
quantifies the impact on forecasts of open-ocean wave heights near an impact site
due to loss of one or more buoys in a network of \DART\ buoys.
The measure is based on a simulation procedure that takes into account
key factors contributing to uncertainty in the forecasts.
One factor is the model for the tsunami signal,
which is never known perfectly in practice,
in part due to uncertainty in the location of the tsunami-generating earthquake.
We mimic this uncertainty
by taking a predetermined central unit source
and relocating it through random selection of a new location for its center.
This relocation leads to perturbations both in the tsunami signal as it appears at \DART\ buoys
and in the open-ocean wave heights.
The perturbations are codified by source coefficients,
which serve as weights for forming the perturbed signal and wave heights
through linear combinations of signals and wave heights
associated with the central unit source and ones abutting it.

Tidal fluctuations and background noise recorded at the buoys
are two additional factors contributing to forecast uncertainty.
We mimic these through the use of archived \DART\ data recorded under ambient conditions
either by the buoy of interest or
by a surrogate operating under similar oceanographic conditions.
The simulated tsunami signal and randomly selected archived data
are added together to form simulated buoy data.
These data are the response in a regression model for estimating source coefficients,
which in turn lead to estimates of wave heights.
Discrepancies between the estimated and presumed coefficients
translate into errors in forecasted wave heights. 
These estimation errors are due in part to tidal fluctuations and background noise
in the simulated buoy data,
but are also influenced by the amount of available data
and by a mismatch between the regression model and the presumed model for the signal,
both of which we mimic in our simulations.

A thousand replications of our simulation procedure for a given unit source/buoy combination
gives a thousand forecasts of open-ocean wave heights at as many impact sites as desired.
To summarize the quality of the forecasts,
we compare the thousand maximum values of the forecasted height
to the corresponding presumed maximum values by forming a root-mean-square error (RMSE).
The RMSEs for a network of buoys in conjunction with a set of unit sources relevant
for a particular impact site form the basis for network evaluation.
The evaluation consists of dropping one or more buoys from the network
and determining what change there is in the minimum RMSE for each unit source.
We propose to use the maximum change in minimum RMSE as a metric for measuring
the health of the buoy network as one or more buoys become inoperative.
We also propose a simple green/yellow/red indicator of network health
as a management tool.

Currently we have bread-boarded our procedure for evaluating the effectiveness
of \DART\ buoy networks in the {\tt R} language
(Ihaka and Gentleman {\bl 1996};
R Development Core Team {\bl 2010}).
Using the {\tt R} code as guidance,
current plans are to encapsulate the methodology we have described into a tool
that would serve multiple purposes.
We would use the tool to go beyond our Aleutian Island and South American case studies
to do a comprehensive study of the existing network of 39 buoys.
This study might identify unit sources for which the current network configuration
is particularly sensitive at certain impact sites to the dropout of buoys. 
We could also use the tool to assess the impact of adjusting the network
by relocating or adding buoys.
Subject to additional research,
it should be possible to use the tool to investigate the use of tide gauges
to help mitigate the loss of buoys.
Despite the simplifications we have made,
we contend that our proposed methodology is an effective and realistic way
to assess network effectiveness
and that a tool built around this methodology would be valuable
for managing networks of \DART\ and other types of tsunami buoys.

\begin{acknowledgements}
This work was funded by the Joint Institute for the Study of the Atmosphere and Ocean (JISAO)
under NOAA Cooperative Agreement No.~NA15OAR4320063 and is JISAO Contribution No.~2714. 
This work is also Contribution No.~4507 from NOAA/Pacific Marine Environmental Laboratory.
The authors thank Peter Dahl for discussion on a running example.
\end{acknowledgements}

\end{document}